\documentclass[aps,prb,amsmath,amssymb,superscriptaddress,floatfix,showpacs,twocolumn]{revtex4-1}
\usepackage{graphicx} 

\usepackage[usenames,dvipsnames]{color}

\usepackage{hyperref}

\begin{document}

\title{The effect of Dzyaloshinskii-Moriya interactions on the phase diagram 
and magnetic excitations of SrCu$_{2}$(BO$_{3}$)$_{2}$}

\author{Judit Romh\'anyi}
\affiliation{Research Institute  for  Solid State  Physics and Optics, H--1525 Budapest, P.O.B.~49, Hungary}
\affiliation{Department of Physics, Budapest University of Technology and Economics, H--1111 Budapest, Budafoki \'ut 8, Hungary}

\author{Keisuke Totsuka}
\affiliation{Yukawa Institute for Theoretical Physics, Kyoto University, Kitashirakawa Oiwake-Cho, Kyoto 606-8502, Japan}

\author{Karlo Penc}
\affiliation{Research Institute  for  Solid State  Physics and Optics, H--1525 Budapest, P.O.B.~49, Hungary}

\date{\today}

\begin{abstract}   
The orthogonal dimer structure in the SrCu$_{2}$(BO$_{3}$)$_{2}$ spin--1/2 magnet provides a realization of the Shastry-Sutherland model. 
Using a dimer--product variational wave function, we map out the phase diagram of the Shastry--Sutherland model including anisotropies. Based on the variational solution, we construct a bond--wave approach to obtain the excitation spectra as a function of magnetic field. The characteristic features of the experimentally measured neutron and ESR spectra are reproduced, like the anisotropy induced zero field splittings 
and the persistent gap at higher fields.
\end{abstract}

\preprint{YITP-10-90}

\maketitle

\section{Introduction}

\begin{figure}[b]
\begin{center}
\includegraphics[width=7truecm]{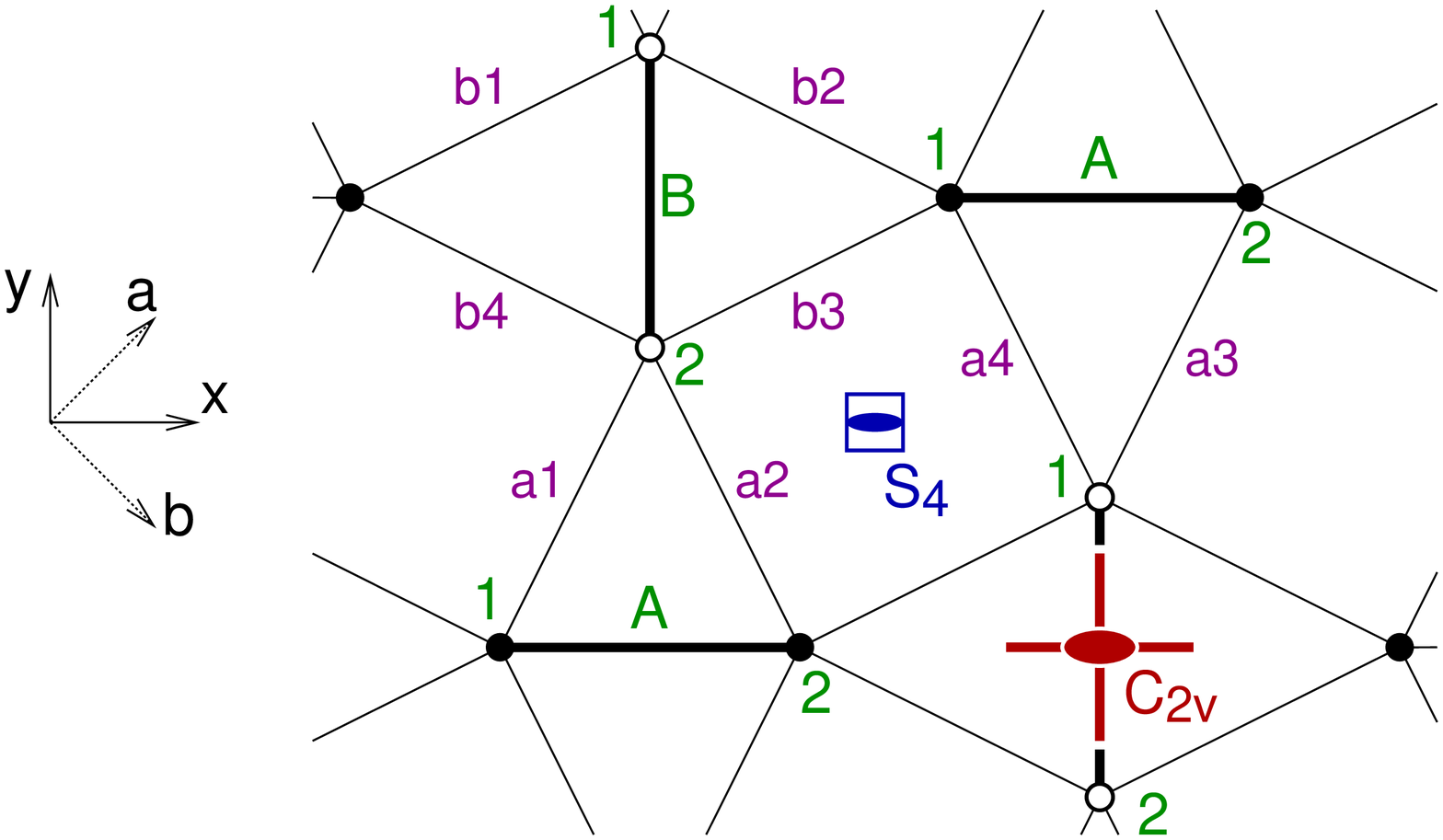}
\caption{(color online) The Shastry--Sutherland lattice in the SrCu$_{2}$(BO$_{3}$)$_{2}$. The orthogonal dimers (thick lines) are denoted by letters A and B, while the Cu ions on the dimer are enumerated as 1 and 2. The inter-dimer bonds (with Heisenberg exchange $J'$) are shown by a thin line. We also show the $\mathcal{S}_4$ and $\mathcal{C}_{2v}$ symmetry point groups of the buckled CuBO$_3$ layer (the open and closed circles indicate the Cu ions that are below or above the layer, respectively).  \label{fig:pointgroups}}
\end{center}
\end{figure}
Spin gap systems, where the spin gap is of quantum mechanical origin,
are of interest to both theoretical and experimental
investigations. These systems have spin-disordered ground states which
can be described as quantum spin liquids\cite{misguich}. 
Spin gap is known to open, for example, in
$S=1/2$ spin systems that form lattices of coupled
dimers\cite{Gelfand-S-H-89}, 
giving rise to many interesting phenomena. 
The Shastry-Sutherland model\cite{Shastry1981} provides 
a unique example of such two-dimensional frustrated networks of 
$S=1/2$ dimers. 
The model includes nearest ($J$) and next nearest neighbor ($J'$) antiferromagnetic interactions as shown in Fig.~\ref{fig:pointgroups}, with the Hamiltonian
\begin{eqnarray}
\mathcal{H}=J\sum_{n.n.}{\bf S}_{i}{\cdot}{\bf S}_{j}
+J'\sum_{n.n.n.}{\bf S}_{i}{\cdot}{\bf S}_{j} \;.
\label{eq:originalmodel}
\end{eqnarray}
In the case of $J'=0$ the model is reduced to a lattice of independent dimers, where in the ground state the $S=1/2$ spins of each dimer form a singlet, and the ground state wave function is just the product of these independent dimer-singlets. According to Shastry and Sutherland\cite{Shastry1981}, the singlet dimer product state is an exact eigenstate of the Hamiltonian even for finite values of the $J'$ due to the particular geometry of the lattice.

 An experimental realization of the Shastry-Sutherland model is the
 quasi two-dimensional antiferromagnetic compound\cite{Kageyama1999-1}
 SrCu$_2$(BO$_3$)$_2$. This material has a tetragonal unit cell and is
 characterized by the alternating layers of CuBO$_3$ molecules and
 Sr$^{2+}$ ions; in the former, the Cu$^{2+}$ ions occupying 
crystallographically equivalent sites carry spin
 $S=1/2$ degrees of freedom and form a lattice of orthogonal dimers 
connected by the triangular shaped BO$_{3}$ molecules%
\cite{Smith1991,Kageyama1999-1}.  
A schematic figure of CuBO$_3$ layer is shown in Fig.~\ref{fig:pointgroups}.

Magnetic susceptibility measurements of this material revealed a peak at
around 20 K and a sharp drop to zero at decreasing
temperatures\cite{Kageyama1999-1}. Fitting the exponential curve that is
characteristic for the spin gap systems, 
Kageyama et al. estimated in Ref.~[\onlinecite{Kageyama1999-1}] 
the gap to be $\Delta\approx 19$ K, while from the NMR 
relaxation rate they obtained the gap about 30 K.  
Magnetization measurements\cite{Kageyama1999-1} clarified the presence of 
a gapped spin-singlet ground state and a continuos transition 
to the gapless magnetic state at 20 T 
which corresponds to a gap of 30 K in a good agreement 
with the relaxation rate measurements. 
While early magnetization measurements in high fields revealed plateaus 
only at 1/4 and 1/8 of the saturated magnetization%
\cite{Kageyama1999-1,Kageyama1999-2}, refined measurements have suggested%
\cite{Onizuka-2000} more plateaus at 1/3 and other values of
magnetization. 

Miyahara and Ueda\cite{Miyahara1999} pointed out that
SrCu$_2$(BO$_3$)$_2$ can be satisfyingly described by the
Shastry-Sutherland model. They estimated the critical point where the
singlet dimer ground state goes to the N\'eel-state to be $(J'/J)_c=0.7$
performing variational calculations and exact numerical diagonalization 
(on the basis of series expansion\cite{Koga2000} and exact diagonalization\cite{Wessel2002}, it is now believed 
that the above two states are mediated by a new plaquette-singlet phase 
and that a transition from the dimer phase to the plaquette-singlet 
occurs at $(J'/J)_c=0.68$).  
Using the experimental findings of Ref.~[\onlinecite{Kageyama1999-1}]
they estimated the nearest neighbor coupling constant to be $J=100$ K 
and the next nearest neighbor coupling $J'=68$ K. This yields
$J'/J=0.68$ indicating SrCu$_2$(BO$_3$)$_2$ to be close 
to the transition point $(J'/J)_c=0.7$.  
Later this estimate has been updated to $J=7.3$ meV with
$J'/J=0.635$ (Ref.~[\onlinecite{Miyahara2000}]) or $J=6.16$ meV with
$J'/J=0.603$ (Ref.~[\onlinecite{Knetter2000}]). 

 Furthermore, Miyahara and Ueda carried out perturbation theory in the dimer state up to the fourth order in $J'/J$ and found that the triplet excitations are localized. The hopping of triplets is only possible through closed paths of dimer bonds, thus only from the sixth order in perturbation. This property of the triplet excitations is related to the formation of plateaus. At certain values of the magnetization the excitations localize into a superlattice structure to minimize the energy\cite{Miyahara1999}.  Momoi and Totsuka\cite{Totsuka2000b,Totsuka2000} 
 have explained the appearance of plateau states through the scenario of metal to Mott-insulator transition where the triplet excitation were treated as bosons interacting via various repulsive interactions 
 arising from higher-order perturbation in $J^{\prime}$. 
 At dominating repulsive interaction, the triplet excitations localize 
 and crystallizes in commensurate patterns developing the plateau states.  In fact, 
 NMR spectroscopy by Kodama et al. exhibited\cite{Kodama2002} directly  the superlattice structure at $m/m_{\text{sat}}=1/8$. 
 Recently new magnetization plateaus have been found\cite{abendschein2008} by nonperturbative 
Contractor--Renormalization (CORE) method at $1/9$, $1/6$ and $2/9$ of the saturation, 
while the analysis\cite{Dorier2008} using the perturbative continuous unitary transformation (PCUT) 
has predicted, on top of the above ones, one more plateau at $2/15$. It has also been argued that that the inclusion of the spin--lattice effects determines the spin structure in the plateaus.\cite{Miyahara2003}

In the past few years various experiments have been carried out to examine the spin excitations. While the original Shastry-Sutherland model is isotropic in spin space, its experimental realization 
SrCu$_2$(BO$_3$)$_2$ exhibits anisotropic behavior; inelastic neutron scattering 
measurements\cite{Cepas2001}, electron spin resonance\cite{Nojiri1999}, 
and Raman scattering\cite{Gozar2005} indicated a splitting of the triplet excitations at the $\Gamma$-point, which was explained to be caused by the effect of inter-dimer Dzyaloshinskii-Moriya (DM) vector directed perpendicular to the copper plane\cite{Cepas2001}. Later another splitting was found\cite{Gaulin2004} 
at the ${\bf q}=(\pi, 0)$ point indicating the relevance of in-plane components of the DM interaction. The ESR study of Nojiri et al.\cite{Nojiri2003} shows an anti-level crossing at the critical magnetic field where 
the lowest-lying triplet excitation would cross the singlet level,   
which is consistent with the persistent spin gap found in the specific-heat\cite{Tsujii2003}- and 
the NMR measurements\cite{Kodama2005}. 
These splittings and the anti-level crossing mean that states corresponding to different magnetization (singlets and triplets) are mixing and $S^z$ is no longer a good quantum number. This mixing between the singlet and triplet states of a dimer can be explained by an intra-dimer anisotropy, e.g. an intra-dimer DM vector. 
For more details on the Shastry-Sutherland model and SrCu$_2$(BO$_3$)$_2$, we refer 
the readers to the review articles Ref.~\onlinecite{Miyahara-U-2003} (theory) and 
Ref.~\onlinecite{Takigawa2010} (experiments).

Magnetization process in such dimer systems as TlCuCl$_3$ 
is fairly well understood; onset of magnetization is triggered by Bose-Einstein 
condensation of gapped triplons and the magnetic phase above the critical field 
is characterized by broken XY-symmetry perpendicular to the applied field 
(see, e.g., Ref.~\onlinecite{Giamarchi-R-T-08} for a review).   Dynamics at high-fields is also 
well described within the above scenario\cite{Matsumoto2004}. 
On the other hand, the existence of DM interactions 
is known to substantially modify the above picture and even new phases may appear 
in the presence of DM interactions.  Moreover, the small kinetic energy and relatively large 
(effective) interactions among dimers lead to various magnetic superstructures
\cite{Kodama2002,Takigawa2008}, which we cannot simply neglect in considering
the dynamics at high fields. 
Nevertheless, the global structure of the phase diagram 
and the dynamics in the presence of DM interactions and magnetic superstructures 
is only partially understood.  The aim of this paper is to present a simple theoretical 
framework to investigate the ground-state phases and the magnetic excitations over them 
with the extension to the cases with superstructures in mind.  
Specifically, by using the bond-wave approximation, 
we examine the excitation spectrum of SrCu$_{2}$(BO$_{3}$)$_{2}$ 
at zero- and low magnetic fields below the plateaus and compare the results with 
the neutron-scattering- and the ESR data.

The paper is structured as follows: in Sec. II we review the symmetry group of SrCu$_{2}$(BO$_{3}$)$_{2}$ 
and determine the allowed anisotropic terms in the Hamiltonian. In Sec. III we describe the variational approach we use to get the ground state and the way we construct the excitation spectrum using the bond operators. 
In Sec. IV, we show the variational phase diagram and the spectrum for the zero field case. In Sec. V we map out the phase diagram in the presence of a field perpendicular to the basal plane.   
In Sections VI and VII we describe the ESR spectra in magnetic fields perpendicular and parallel to the plane, respectively.
Last, we compare our results with neutron scattering experiments and the ESR spectra in Sec. VIII.
We conclude with Sec. IX.

\section{Symmetry considerations and the model Hamiltonian}

At high temperatures the space group of SrCu$_{2}$(BO$_{3}$)$_{2}$ is a tetragonal $I4/mcm$ (Refs.~[\onlinecite{Smith1991}] and [\onlinecite{sparta2001}]). 
A structural distortion in the CuBO$_{3}$ layers below $T_\text{s}=395$ K shifts the two types of orthogonal dimer planes along the $z$ axis in opposite direction, lowering the symmetry of the 
SrCu$_{2}$(BO$_{3}$)$_{2}$ to $I\bar42m$ at low temperatures, also with tetragonal symmetry\cite{sparta2001}. Restricting ourselves to the symmetries of the CuBO$_3$ layer, above $T_{s}$ the two --- to each other orthogonal --- types of dimers lay in the same plane and the wallpaper group 
$p4g$ consists of the point group  
$\mathcal{C}_{2v}=\left\{E, C_{2}(z), \sigma_{xz}, \sigma_{yz}\right\}$ at the middle of dimer bond and of the point group 
$\mathcal{C}_{4h}$ in the middle of four sites belonging to different dimers. Lowering the temperature, while 
the $\mathcal{C}_{2v}$ remains a symmetry of the buckled CuBO$_3$ layers, the loss of the $\sigma_h$ reflection plane below $T_\text{s}$ lowers the $\mathcal{C}_{4h}$ point group to $\mathcal{S}_4=\left\{E, S_{4}, C_{2}(z), S^3_{4}\right\}$ (see Fig.~\ref{fig:pointgroups}).The unit cell in both cases consists of two orthogonal dimers --- dimer A that is parallel to the $x$ axis, and dimer B parallel to $y$ axis.  
\begin{figure}[t]
\begin{center}
\includegraphics[width=5truecm]{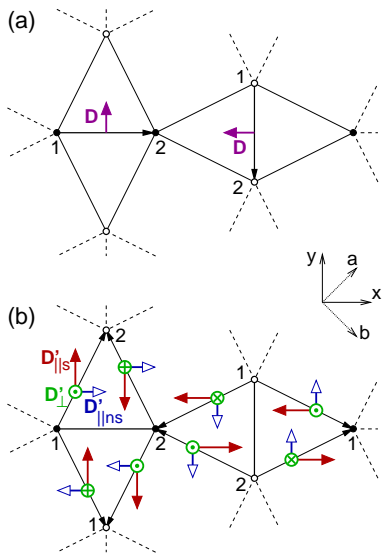}
\caption{(color online) The components of the symmetry allowed intra-dimer (a) and inter-dimer (b) Dzyaloshinskii-Moriya vectors shown in a unit cell. 
The arrows running from $i$ to $j$ 
on the bonds indicate the ordering of the spin operators 
$\mathbf{S}_{i}{\times}\mathbf{S}_{j}$ in the DM interaction term.
\label{fig:dm}}
\end{center}
\end{figure}

The symmetry of the lattice determines the possible terms in the Hamiltonian: the components of the $g$-tensor anisotropy and of the exchange interactions, including the components of the DM interactions; see Table~\ref{table:correspondence} and, e.g., Ref.~[\onlinecite{Kodama2005}]. Here we shortly present the relevant terms. The sites A1, A2, B1, and B2 correspond to the sites 1, 2, 3, and 4 of Ref.~[\onlinecite{Kodama2005}], respectively.

 The $g$-tensor at site A1 takes the following form:
 \begin{equation}
{\bf g}_{\text{A1}} = \left(
\begin{array}{ccc}
 g_x & 0  & -g_{\text{s}}  \\
 0 & g_y  &  0 \\
 -g_{\text{s}} & 0  &  g_z 
\end{array}
\right) 
\label{eq:gtensor}
 \end{equation} 
 with $g_{x}=g_{y}$ as required by the tetragonal symmetry. 
 When the field is in the $z$ direction, the Zeeman term reads
 \begin{subequations}
\begin{equation}
\begin{split}
\mathcal{H}_{h} =& - g_z \mu_\text{B} H_z \left( S^{z}_{\text{A1}}+S^{z}_{\text{A2}}
+S^{z}_{\text{B1}}+S^{z}_{\text{B2}} \right)  \\
  & + g_{\text{s}} \mu_\text{B} H_z \left( S^{x}_{\text{A1}}-S^{x}_{\text{A2}}
  +S^{y}_{\text{B1}}-S^{y}_{\text{B2}} \right) \;,
  \end{split}
\end{equation}
while if the field is along the $x$ axis, it reads 
\begin{equation}
\begin{split}
  \mathcal{H}_{h} =& 
      - g_x \mu_B H_x\left(S^{x}_{\text{A1}}+S^{x}_{\text{A2}}\right) 
      - g_y H_x \left(S^{x}_{\text{B1}}+S^{x}_{\text{B2}}\right) \\
      & + g_{\text{s}} \mu_\text{B} H_x\left( S^{z}_{\text{A1}}-S^{z}_{\text{A2}} \right) \;.
\end{split}
\end{equation}
\end{subequations}
For convenience, we choose $h_z = g_z \mu_B H_z$ and introduce the scaled variable
$\tilde g_{\text{s}} = g_{\text{s}}/g_z$.

Next, we consider the DM interactions $\mathcal{H_{\rm DM}}= \mathcal{H}_D+\mathcal{H}_{D'}$, with the intra-dimer
\begin{equation}
\mathcal{H}_D = 
 \sum_{\text{NN}} \mathbf{D}_{ij}{\cdot} \left(\mathbf{S}_{i}\times\mathbf{S}_{j}\right)
\label{eq:HD}
\end{equation}
and inter-dimer
\begin{equation}
\mathcal{H}_{D'} = 
\sum_{\text{NNN}} \mathbf{D'}_{ij} {\cdot} \left(\mathbf{S}_{i}\times\mathbf{S}_{j}\right)  \\
\label{eq:HDp}
\end{equation}
contributions. The summation is over the nearest neighbor (NN) and next-nearest neighbor (NNN) sites.
Once we specify the DM vectors on a bond, the DM interactions on the remaining bonds of the unit cell follow from the symmetry of the cell, as shown in Fig.~\ref{fig:dm}. To specify the sign of the DM interaction unambiguously, we have also denoted by an arrow the order of the spins in the cross product: an arrow from site $i$ to site $j$ in the figure means that we need to take $\mathbf{S}_{i}\times\mathbf{S}_{j}$. 

At temperatures above the structural transition\cite{sparta2001} 
$T>T_{\text{s}}=395$ K (high-temperature phase), 
the middle of a bond is an inversion center due to which there is no DM vector 
on the dimer bond and only the inter-dimer ${\bf D'}$ perpendicular to 
the CuBO$_3$ plane is allowed:
\begin{subequations}
\begin{align}
 & {\bf D}=0 \;,\\
 & {\bf D'} = (0,0,D'_{\bot}) 
\end{align}
\end{subequations}
 (we denote by $D'_{\bot}$ the $z$ component).  
 In the following, we will call this case {\em the high-symmetry case}.
  
Below $T_{\text{s}}$ (low temperature phase), however, this inversion symmetry 
 is lost and the in-plane DM components are allowed as well. 
Correspondingly, ${\bf D'}$ becomes an arbitrary vector 
and the intra-dimer ${\bf D}$ is lying in the CuBO$_3$-plane and perpendicular 
to the dimer, so that 
\begin{subequations}
\begin{align}
& {\bf D}_{\text{A}}=(0,D,0) \;,\\
& {\bf D}_{\text{B}}=(-D,0,0) \;,
\end{align}
\end{subequations}
with the site ordering convention as shown in Fig.~\ref{fig:dm}.
We refer to this case as {\em the low-symmetry case}. 

A discussion of the different estimates for the strength of the DM interactions 
and the $g$-tensor anisotropies for SrCu$_{2}$(BO$_{3}$)$_{2}$ is presented 
in Sec.~\ref{sec:exp}.   
\begin{table}
\caption{Summary of symmetry analyses.  If the $g$-tensor anisotropy is taken 
into account, the spin $O(2)$-symmetry is lost.%
\label{table:correspondence}
}
\begin{center}
\begin{ruledtabular}
\begin{tabular}{lcc}
\raisebox{0.5ex}[0pt]{} &
\raisebox{0.5ex}[0pt]{high-symmetry} & 
\raisebox{0.5ex}[0pt]{low-symmetry} \\ 
\hline
symmetry (unit cell) & ${\cal D}_{4h}$ & ${\cal D}_{2d}$  \\ 
DM (intra-dimer) & forbidden & $\mathbf{D}\parallel (ab) \wedge \mathbf{D}\perp$ dimer \\   
DM (inter-dimer) & $ \mathbf{D}^{\prime}\parallel c $ & arbitrary  \\ 
spin ($h \parallel z$) & $O(2)$-sym & $-$  \\   
spin ($h \parallel x$) & $-$ & $-$  
\end{tabular}
\end{ruledtabular}
\end{center}
\end{table}
\section{The variational approach and the bond--wave theory}
\label{sec:method}
\subsection{Variational wave function}
\label{sec:varwf}
 The ground state of the pure Shastry-Sutherland model (\ref{eq:originalmodel}) 
 can be written as a product of singlets $|s\rangle$ over the dimer bonds, $\Psi = \prod_{\text{dimers}} |s\rangle$. In the presence of the DM interactions and finite magnetic fields, 
we need to extend this wave functions to a variational one. Namely, we allow for a linear combination of the singlet and triplet states on each dimer (we keep the dimer wave function entangled), while we retain the product form over the dimer bonds:
\begin{equation}
 |\Psi\rangle = \prod_{\text{A dimers}} |\psi_\text{A}\rangle \prod_{\text{B dimers}} |\psi_\text{B}\rangle \;,
\label{eqn:var-ansatz}
\end{equation}
where 
\begin{subequations}
\label{eq:varpsiAB}
\begin{eqnarray}
| \psi_\text{A} \rangle &=& u_s |s\rangle + \sum_\alpha u_\alpha |t_\alpha\rangle \;,
\\
| \psi_\text{B} \rangle &=& v_s |s\rangle + \sum_\alpha v_\alpha |t_\alpha\rangle \;,
\label{eq:varpsiABB}
\end{eqnarray}
\end{subequations}
with $|t_\alpha \rangle$ being the three components of the triplets. This wave function can describe the phases that do not break the translational symmetry.  
Since we have two (i.e. A and B) dimers in the unit cell, 
the entire wave function $|\Psi\rangle$ is translationally invariant even when the wave functions 
of the two dimers are different. Certainly, this wave function cannot describe the plateaus except for the one at 1/2; for other values one would need to take a larger unit cell. The variational parameters $u$ and $v$ are then determined by minimizing the energy
\begin{equation}
  E = \frac{\langle \Psi | \mathcal{H} | \Psi \rangle}{\langle \Psi | \Psi \rangle} \;.
  \label{eqn:var-energy}
\end{equation}
The minimization is performed numerically, except for some simple cases when we could find analytical solutions. 

\subsection{Auxiliary boson formalism for the Hamiltonian}
\label{sec:aux-boson}
In order to find the excitation spectrum, we introduce, in the spirit of Sachdev 
and Bhatt\cite{sachdev}, auxiliary bosons which create the singlet and the triplet  
on each bond;  the operator $s^\dagger$ creates the singlet state $(|\uparrow\downarrow\rangle - |\downarrow \uparrow \rangle)/\sqrt{2} $, while the operators $t^{\dagger}_{x}$, $t^{\dagger}_{y}$, and $t^{\dagger}_{z}$ create the triplet states 
$i(|\uparrow\uparrow\rangle - |\downarrow \downarrow \rangle)/\sqrt{2}$, $(|\uparrow\uparrow\rangle + |\downarrow \downarrow \rangle)/\sqrt{2}$, and $-i(|\uparrow\downarrow\rangle + |\downarrow \uparrow \rangle)/\sqrt{2}$, respectively.  
This definition is different from the one used in Ref.~[\onlinecite{sachdev}] 
by an additional phase factor $-i$ which ensures that the new bosons are time-reversal invariant%
\footnote{The action of the antiunitary time-reversal operator on spin-1/2 states 
flips the spin and adds minus sign only to $(-1/2)$-state. 
This operation leaves the singlet- and the triplet operators  
$s^{\dagger}$, $t^{\dagger}_{x}$, $t^{\dagger}_{y}$, and $t^{\dagger}_{z}$ 
defined in section \ref {sec:aux-boson} invariant.
}.  
Furthermore, in order for the four states on each dimer to be faithfully represented, 
the number of bosons per dimer is constrained: 
\begin{align}
s^{\dagger}s^{\phantom{\dagger}}+\sum_{\alpha=x,y,z}t^{\dagger}_{\alpha}t^{\phantom{\dagger}}_{\alpha}=1 \;.
\label{eq:cond}
\end{align}
The components of the spin operator at bond $j$ are then given as
\begin{subequations}
\label{eq:Scomp}
\begin{eqnarray}
S^{\alpha}_{j,1} &=&
 \frac{i}{2} 
\left( 
t^{\dagger}_{\alpha,j} s^{\phantom\dagger}_{j} 
- 
s^{\dagger}_{j} t^{\phantom\dagger}_{\alpha,j} 
\right)
- \frac{i}{2} \epsilon_{\alpha,\beta,\gamma} t^{\dagger}_{\beta,j} t^{\phantom\dagger}_{\gamma,j} 
 \;,\\
S^{\alpha}_{j,2} &=& -\frac{i}{2}
\left(
t^{\dagger}_{\alpha,j}s^{\phantom\dagger}_{j}
- 
s^{\dagger}_{j}t^{\phantom\dagger}_{\alpha,j}
\right)
-\frac{i}{2} \epsilon_{\alpha,\beta,\gamma} t^{\dagger}_{\beta,j} t^{\phantom\dagger}_{\gamma,j}
 \;.
\end{eqnarray}
\end{subequations}

The intra-dimer part of the Heisenberg Hamiltonian $\mathcal{H}$ [Eq.~(\ref{eq:originalmodel})] in the bond representation reads 
\begin{equation}
\mathcal{H}_{J} = - \frac{3 J}{4} \sum_{j}s^{\dagger}_{j}s^{\phantom\dagger}_{j}+\frac{J}{4}\sum_{j}\sum_{\alpha=x,y,z}t^{\dagger}_{\alpha,j}t^{\phantom\dagger}_{\alpha,j} \;,
\end{equation}
while the intra-dimer DM-interaction reads
\begin{eqnarray}
\mathcal{H}_{D}&=&\frac{D}{2}\sum_{j \in A}\left(t^{\dagger}_{y,j}s^{\phantom\dagger}_{j} + s^{\dagger}_{j}t^{\phantom\dagger}_{y,j}\right) 
\nonumber\\ &&
-\frac{D}{2}\sum_{j\in B}\left(t^{\dagger}_{x,j}s^{\phantom\dagger}_{j} + s^{\dagger}_{j}t^{\phantom\dagger}_{x,j}\right) \;.
\label{eq:intra}
\end{eqnarray}
We can derive similar expressions for all the terms in the Hamiltonian. 

In the presence of magnetic field along the $z$ axis, it is more convenient to use 
the triplet bosons $t^{\dagger}_{1}$, $t^{\dagger}_{0}=it^{\dagger}_{z}$, 
and $t^{\dagger}_{\bar 1}$ that create $|\uparrow\uparrow\rangle$, $(|\uparrow\downarrow\rangle + |\downarrow \uparrow \rangle)/\sqrt{2}$, and 
$|\downarrow \downarrow \rangle$ that are the eigenstates of the $z$-component of the spin operator. The spin operators then read:
\begin{subequations}
\label{eq:Scompu0d}
\begin{eqnarray}
S^{+}_{j,l}&=& \frac{t^{\dagger}_{1,j} t^{\phantom{\dagger}}_{0,j} + t^{\dagger}_{0,j} t^{\phantom{\dagger}}_{\bar 1,j}}{\sqrt{2}}  \pm \frac{s^{\dagger}_j t^{\phantom{\dagger}}_{\bar 1,j} - t^{\dagger}_{1,j} s^{\phantom{\dagger}}_j}{\sqrt{2}} \;, \\
S^{-}_{j,l}&=& \frac{t^{\dagger}_{\bar 1,j} t^{\phantom{\dagger}}_{0,j} + t^{\dagger}_{0,j} t^{\phantom{\dagger}}_{1,j}}{\sqrt{2}}  \mp \frac{s^{\dagger}_j t^{\phantom{\dagger}}_{1,j} - t^{\dagger}_{\bar 1,j} s^{\phantom{\dagger}}_j}{\sqrt{2}} \;, \\
S^{z}_{j,l}&=& \frac{t^{\dagger}_{1,j} t^{\phantom{\dagger}}_{1,j} - t^{\dagger}_{\bar 1,j} t^{\phantom{\dagger}}_{\bar 1,j}}{2}  \pm \frac{s^{\dagger}_j t^{\phantom{\dagger}}_{0,j} + t^{\dagger}_{0,j} s^{\phantom{\dagger}}_j}{2} \;,
\end{eqnarray}
\end{subequations}
where the upper sign is for $l=1$ spin and the lower sign for the $l=2$ spin in the dimer, as denoted in Fig.~\ref{fig:pointgroups}.

\subsection{Bond wave method}
\label{sec:BondWave}
After rewriting the Hamiltonian in terms of the bond operators using Eq.~(\ref{eq:Scomp}) or (\ref{eq:Scompu0d}), we perform a bond-wave approximation which is a natural extension 
of the usual spin wave theory. In the bond-wave approximation we extend the number of bosons per dimer from 1 to $M$, so that the constraint (\ref{eq:cond}) now reads
\begin{equation}
s^{\dagger}s^{\phantom{\dagger}}+\sum_{\alpha=x,y,z}t^{\dagger}_{\alpha}t^{\phantom{\dagger}}_{\alpha}=M \;.
 \label{eq:condM}
\end{equation}
The variational approach mentioned in Sec.~\ref{sec:varwf} is analogous to finding 
the classical ($S\rightarrow\infty$) ground state for spin models; $M\rightarrow\infty$ is the classical solution where the quantum fluctuations between the dimers are neglected. 
To see this, it is convenient to rotate the `quantization axis'.   
Then, the variational solution $| \psi_\text{A} \rangle$ in Eq.~(\ref{eq:varpsiAB}) can be 
written in terms of the `rotated' bosons as $| \psi_\text{A} \rangle 
= \tilde s^\dagger_\text{A} |0\rangle$, where the $\tilde s^\dagger_\text{A} = u_s s^\dagger + \sum_\alpha u_\alpha t_\alpha^\dagger$. Similarly, we have $| \psi_\text{B} \rangle = \tilde s^\dagger_\text{B} |0\rangle$ with $\tilde s^\dagger_\text{B}$ defined using Eq.~(\ref{eq:varpsiABB}).  
In the case of general $M$, we promote the above expressions 
to $| \psi_\text{A} \rangle = (\tilde s^\dagger_\text{A})^M |0\rangle$ and 
$| \psi_\text{B} \rangle = (\tilde s^\dagger_\text{B})^M |0\rangle$, which are direct analogues of 
the Bloch coherent states for the spin-$M/2$ system.  In the classical-limit $M\rightarrow \infty$, the coherent state 
$|\psi_{\text{A,B}}\rangle$ may be thought of as the condensate of $\tilde{s}_{\text{A,B}}$. 
We also rotate the remaining bosons into $\tilde t_\alpha$ so that they obey the usual commutation relations and the local constraint Eq.~(\ref{eq:condM}).  
Accordingly, the expressions of the spin operators (\ref{eq:Scompu0d}) get modified. 

To consider the small `transverse' fluctuations around the classical solution,  
we solve the constraint explicitly for $\tilde{s}$ and treat $\tilde{t}$s 
as the Holstein-Primakoff bosons. 
Using the formal expansion (valid to order shown)
\begin{eqnarray}
  \tilde s^\dagger = \tilde s^{\phantom{\dagger}} &=& 
    \sqrt{M - \sum_\alpha \tilde t_\alpha^\dagger  \tilde t_\alpha^{\phantom{\dagger}}}
     \nonumber\\
  & \approx & \sqrt{M} - 
  \frac{1}{\sqrt{M}}
  \sum_\alpha \tilde t_\alpha^\dagger \tilde t_\alpha^{\phantom{\dagger}} + \cdots\;,
\end{eqnarray}
we perform a $1/M$-expansion in the spin operators and subsequently in the Hamiltonian.  Then the Hamilton operator can be written as 
\begin{equation}
 \mathcal{H} = M^2 \mathcal{H}^{(0)} + M^{3/2} \mathcal{H}^{(1)} + M \mathcal{H}^{(2)} + \cdots \label{eq:Ham_bwexp} 
\end{equation}
where $\mathcal{H}^{(0)}=E_0$ is just the variational energy and $\mathcal{H}^{(1)}$ is 
a collection of the terms that are linear in 
\begin{equation}
\mathbf{\tilde t}^{\phantom{\dagger}}_\mathbf{k} = \left(
\tilde t^{\phantom{\dagger}}_{x,\text{A},\mathbf{k}},
\tilde t^{\phantom{\dagger}}_{y,\text{A},\mathbf{k}},
\tilde t^{\phantom{\dagger}}_{z,\text{A},\mathbf{k}},
\tilde t^{\phantom{\dagger}}_{x,\text{B},\mathbf{k}},
\tilde t^{\phantom{\dagger}}_{y,\text{B},\mathbf{k}},
\tilde t^{\phantom{\dagger}}_{z,\text{B},\mathbf{k}} \right)
\end{equation}
 and in the similarly defined
 $\mathbf{\tilde t}^{\dagger}_\mathbf{k}$. 
As is expected from the variational nature, it turns out that $\mathcal{H}^{(1)}$ is identically equal to 0 for the variational solution. The quadratic part $\mathcal{H}^{(2)}$ is of the form
\begin{eqnarray}
\mathcal{H}^{(2)}&=& \frac{1}{2} \sum_{\mathbf{k}\in \text{BZ}}
 \left(\begin{array}{c}
\mathbf{\tilde t}^{\dagger}_{\bf k}\\
\mathbf{\tilde t}^{\phantom{\dagger}}_\mathbf{-k}
 \end{array}\right)^T
\left(\begin{array}{cc}
M&N\\
N^*&M\\
\end{array}\right)
\left(\begin{array}{c}
\mathbf{\tilde t}^{\phantom{\dagger}}_\mathbf{k}\\
\mathbf{\tilde t}^{\dagger}_{\bf -k}
\end{array}\right) \;.
\end{eqnarray}
The $\mathcal{H}^{(2)}$ can be diagonalized by the method described in Appendix, 
Sec.~\ref{sec:Appendix_bondwave}, and we find three excitations (one for each $\tilde t_\alpha$ boson) per dimer.

\section{Bond--wave spectrum in zero field}
\label{sec:Dfin_h0}
We start with the discussion of the zero-field excitation spectra in the low-symmetry (finite $D$) case. 
Early neutron scattering results\cite{Kageyama2000} indicated that the spectrum consists of essentially dispersionless (localized) single-triplet branch and other multi-triplet 
ones. 
This is the consequence of the orthogonal dimer structure, and triplets get dispersion in the 6$^\text{th}$ order of the perturbation expansion \cite{Miyahara1999,Miyahara-U-2003} in $J'/J$. Later, higher-resolution neutron scattering 
experiments\cite{Gaulin2004} revealed that the first triplet excitation actually splits into 3 subbands with well-defined dispersions. The splitting indicates the presence of anisotropies. In the following, we will calculate these spectra starting from the bond-wave theory.

 The variational wave function that minimizes the energy (\ref{eqn:var-energy}) 
 in zero magnetic field takes the following form:
\begin{subequations}
 \label{eq:WFh0I}
\begin{eqnarray}
 | \psi_\text{A} \rangle &\propto& |s\rangle + \frac{w}{\sqrt{2}} \left(|t_{\bar 1}\rangle  + |t_{1}\rangle \right)  = |s\rangle + w |t_y\rangle \;,  \label{eq:WFh0Ia}
 \\
 | \psi_\text{B} \rangle &\propto& |s\rangle + \frac{w}{\sqrt{2}} \left(i|t_{\bar 1}\rangle -i|t_{1}\rangle  \right )  = |s\rangle - w |t_x\rangle \; ,
\end{eqnarray}
\end{subequations}
with 
\begin{equation}
w=-\frac{D}{J+\sqrt{J^2+D^2}} = -\frac{D}{2J} + O(D^3/J^3) \; .
\label{eq:wIh0}
\end{equation}
The corresponding energy is given by
\begin{equation}
E_{Z_1[\mathcal{D}_{2d}]}=-\frac{J}{2}-\sqrt{D^2+J^2} \;.
  \label{eq:w_ho}
\end{equation}
This wave function is time-reversal invariant and it does not break any of the symmetries of the $\mathcal{D}_{2d}$, the plane group of the Hamiltonian. We denote this phase by $Z_1[\mathcal{D}_{2d}]$.

To get the excitation spectrum following the recipe outlined in Sec.~\ref{sec:BondWave}, we rotate the states on each bond of type A as
\begin{eqnarray}
\left(\begin{array}{c}
\tilde{s}^{\dagger}_{\text{A}} \\
\tilde{t}^{\dagger}_{x,\text{A}} \\
\tilde{t}^{\dagger}_{y,\text{A}} \\
\tilde{t}^{\dagger}_{z,\text{A}} 
\end{array}\right) =
\left(\begin{array}{cccc}
\frac{1}{\sqrt{1+w^2}} & 0 & \frac{w}{\sqrt{1+w^2}} & 0\\
0 & 1 & 0 & 0\\
 -\frac{w}{\sqrt{1+w^2}} & 0 &  \frac{1}{\sqrt{1+w^2}} & 0\\
0 & 0 & 0 & 1
\end{array}\right)
\left(\begin{array}{c}
s^{\dagger}_{\text{A}} \\
t^{\dagger}_{x,\text{A}} \\
t^{\dagger}_{y,\text{A}} \\
t^{\dagger}_{z,\text{A}} 
\end{array}\right) \;,
\end{eqnarray} 
with an analogous rotation on bonds B,
so that the variational wave functions in Eqs.~(\ref{eq:WFh0I}) are given as $| \psi_\text{A} \rangle = \tilde{s}^{\dagger}_{\text{A}} |0\rangle$ and $| \psi_\text{B} \rangle = \tilde{s}^{\dagger}_{\text{B}} |0\rangle$. Next, we condense the $\tilde{s}^{\dagger}_{\text{A}}$ and  $\tilde{s}^{\dagger}_{\text{B}}$ singlets. The expression of the bond-wave Hamiltonian is complicated for arbitrary point in the Brillouin-zone, except at the $\Gamma$ point, where it assumes the following form: \begin{widetext}
\begin{eqnarray}
\mathcal{H}^{(2)}&=&
\frac{\Omega}{2} \left(\tilde{t}^{\dagger}_{z,\text{A}} \tilde{t}^{\phantom{\dagger}}_{z,\text{A}} +\tilde{t}^{\dagger}_{z,\text{B}} \tilde{t}^{\phantom{\dagger}}_{z,\text{B}} +\tilde{t}^{\phantom{\dagger}}_{z,\text{A}} \tilde{t}^{\dagger}_{z,\text{A}} +\tilde{t}^{\phantom{\dagger}}_{z,\text{B}} \tilde{t}^{\dagger}_{z,\text{B}}   \right)\nonumber\\
&+& \frac{1}{2}  
\left(\begin{array}{c}
\tilde{t}^{\dagger}_{x,\text{B}}  \\
\tilde{t}^{\dagger}_{y,\text{A}} \\
\tilde{t}^{\phantom{\dagger}}_{y,\text{A}} \\
\tilde{t}^{\phantom{\dagger}}_{x,\text{B}}  
\end{array}\right)^T
\left(\begin{array}{cccc}
\sqrt{J^2+D^2} & -2D'_\perp & -2D'_\perp & 0\\
-2D'_\perp & \sqrt{J^2+D^2} & 0 & -2D'_\perp\\
-2D'_\perp & 0 & \sqrt{J^2+D^2} & -2D'_\perp\\
0 & -2D'_\perp & -2D'_\perp & \sqrt{J^2+D^2} 
\end{array}\right)
\left(\begin{array}{c}
\tilde{t}^{\phantom{\dagger}}_{x,\text{B}}  \\
\tilde{t}^{\phantom{\dagger}}_{y,\text{A}}  \\
\tilde{t}^{\dagger}_{y,\text{A}} \\
\tilde{t}^{\dagger}_{x,\text{B}}  
\end{array}\right)\nonumber\\
&+& \frac{1}{2} 
\left(\begin{array}{c}
\tilde{t}^{\dagger}_{y,\text{B}}  \\
\tilde{t}^{\dagger}_{x,\text{A}} \\
\tilde{t}^{\phantom{\dagger}}_{x,\text{A}} \\
\tilde{t}^{\phantom{\dagger}}_{y,\text{B}}  
\end{array}\right)^T
\left(\begin{array}{cccc}
\Omega & \Xi & \Xi & 0\\
\Xi & \Omega & 0 & \Xi\\
\Xi & 0 & \Omega & \Xi\\
0 & \Xi & \Xi & \Omega 
\end{array}\right)
\left(\begin{array}{c}
\tilde{t}^{\phantom{\dagger}}_{y,\text{B}}  \\
\tilde{t}^{\phantom{\dagger}}_{x,\text{A}}  \\
\tilde{t}^{\dagger}_{x,\text{A}} \\
\tilde{t}^{\dagger}_{y,\text{B}}  
\end{array}\right) \;,
 \label{eq:Hamh0Z1}
\end{eqnarray}
\end{widetext}
with
\begin{eqnarray}
\Xi &=& \frac{2(D'_\perp+wD'_{||}+J'w^2)}{1+w^2} \;, \\
\Omega &=& \frac{J+\sqrt{J^2+D^2}}{2} \;.
\end{eqnarray}
For simplicity, we have introduced the quantity
\begin{equation}
D'_{||} = D'_{||,\text{ns}}- D'_{||,\text{s}} \;,
\end{equation}
for the in-plane components of the inter-dimer DM interaction 
as only this combination enters the variational ground state energy of the translationally invariant dimer-product wave function (\ref{eqn:var-ansatz}) and excitations. Actually, the matrices in the Hamiltonian (\ref{eq:Hamh0Z1}) can be reduced to $2\times 2$ ones by using the symmetries of the $\mathcal{S}_4$ point group (see Eq.~\ref{eq:tsympm}). 

 Following Appendix \ref{sec:appendix_general}, we diagonalize the Hamiltonian (\ref{eq:Hamh0Z1}) and get the excitation energies: 
\begin{subequations}
\label{eq:wh0}
\begin{align}
& \omega_{1,2} = \Omega  \;,\\
& \omega^\pm_3 = \sqrt{\Omega(\Omega \pm 2 \Xi)} \;,
  \\
& \omega^\pm_4 = \sqrt{J^2+D^2 \pm 4 D'_\perp \sqrt{J^2+D^2}} \;.
\end{align}
\end{subequations}

We note that $\omega^\pm_3 \approx J \pm 2 D'_{\bot}$ and $\omega^\pm_4 \approx J \pm 2 D'_{\bot}$ for small values of $D'_{\bot}/J$, thus the pairs of excitations 
$\omega^{a}_{3}$ and $\omega^{a}_{4}$ ($a=\pm$) are essentially indistinguishable.  
Furthermore, the splitting between the two branches 
$\omega^-_{n}$ and $\omega^+_{n}$ ($n=3,4$) at the $\Gamma$ point is 
$4 D'_{\bot}+O({D'_\perp}^2/J)$, which will be used to estimate the value of $D'_{\bot}$ 
in Sec.~\ref{sec:qualitative-comp}.  

A similar calculation at the $\mathbf{q}=(\pi,\pi)$ point gives 
\begin{eqnarray}
\omega_{1} &=& \frac{J}{\sqrt{J^2+D^2}} \;, \\
\omega_{2} &=& \frac{J}{2}+\frac{J^2}{2\sqrt{J^2+D^2}} \;, 
\end{eqnarray}
and splittings that are quadratic in DM interactions:
\begin{eqnarray}
 \omega_{2}-\omega_{1} = \frac{J}{2}-\frac{J^2}{2\sqrt{J^2+D^2}} \approx \frac{D^2}{4J^2} \;.
\end{eqnarray}
 
 Let us note that for larger values of $|D'_\bot|$ the dispersion becomes comparable to the gap, and new phases appear. The branches $\omega_3^\pm$ become gapless when 
 $\sqrt{J^2+D^2} = \mp 4 (D'_\perp + w D'_{||}+J'w^2)$, while 
$\omega_4^\pm=0$ for $4D'_\perp=\mp \sqrt{J^2+D^2}$. Assuming that $D'_\|$ is absent and keeping only the leading term in $D/J$, we get that the phase $Z_1[\mathcal{D}_{2d}]$ 
is stable for 
\begin{equation}
-\frac{J}{4} - \frac{D^2}{8 J} <D'_\perp < \frac{J}{4} - \frac{D^2}{8 J^2} \left(2J'-J\right) \;.
\end{equation}
in the zero field. Beyond these boundaries $Z_2$ twofold degenerate phases with the symmetry group $\mathcal{C}_{2v}$ (when $\omega_3^-\rightarrow0$ for $D'_\perp>0$) or $\mathcal{S}_4$ (when $\omega_4^+\rightarrow0$ for $D'_\perp<0$) is realized (see also Fig.~\ref{fig:phaseD01_hz}). We will discuss these $Z_2$ phases in more detail in a separate publication.

\begin{figure}[h]
\begin{center}
\includegraphics[width=8 truecm]{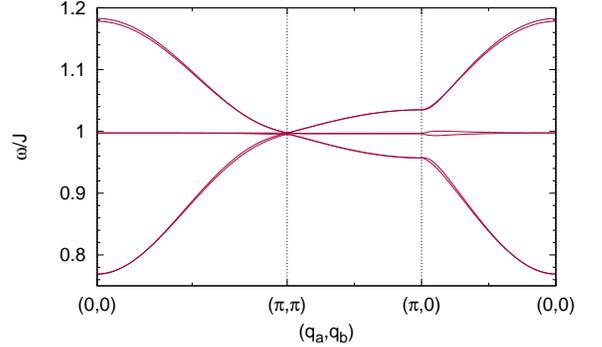}
\caption{(color online) Dispersions of the quasi--triplet excitations in zero magnetic field: 
$\omega_{3,4}^{-}$ (bottom), $\omega_{1,2}$ (middle), and $\omega_{3,4}^{+}$ (top). 
We have chosen $D=D'_\perp=0.1J$ and $J'=0.6 J$.}
\label{fig:swdisp}
\end{center}
\end{figure}

 After a first order expansion with respect to $D/J$, $D'_{\bot}/J$, $D'_{||,s}/J$, and $D'_{||,ns}/J$, it is possible to solve the eigenvalue problem analytically. In this case we get three two-fold degenerate branches: a dispersionless one with eigenvalue $J$ and two branches with 
\begin{equation}
\omega_{\textbf{q}}=\sqrt{J^2\pm J\Omega_{\textbf{q}}}\approx J\pm\frac{1}{2}\Omega_{\textbf{q}}\;,
\end{equation}
where
\begin{eqnarray}
\Omega_{\textbf{q}}=\left[\left(\frac{J' D}{J}-2 D'_{\|,s}\right)^2(1-\cos{q_a}\cos{q_b})\right.+\nonumber\\
\left.+16 D'^2_{\bot}\cos^2{\frac{q_a}{2}}\cos^2{\frac{q_b}{2}}\right]^{1/2} \;.
\label{eq:spect}
\end{eqnarray}
The dispersion of the quasi triplet excitations is shown in Fig. \ref{fig:swdisp}.

The dispersion in zero field has also been considered by Cheng {\it et al.} in Ref.~[\onlinecite{Cheng2007}], 
where they used a different approach; by suitable rotation of the spin operators they removed the $D$ and arrived at an effective Hamiltonian, where they carried out a first order perturbation expansion to get the dispersions of the effective triplets. Even though they considered a unit cell that has $\mathcal{C}_4$ symmetry, our dispersion agrees with their result, up to ambiguity in the sign in front of the $D'_{||,s}$ in Eq.~(\ref{eq:spect}). Furthermore, they extended their analysis by exact diagonalization calculations of the spectra.  

\section{Phase diagram in a field parallel to $z$ axis}
In this section, we are going to consider the variational ground-state phase diagram 
in the presence of an external field along the $z$-axis.  
The full Hamiltonian is now invariant under the magnetic group 
$\mathcal{S}_{4}+\Theta\sigma_{xz}{\times}\mathcal{S}_{4}$ which is isomorphic to 
$\mathcal{D}_{2d}$.  For clarity of the argument, we investigate 
the high-symmetry case (where $D=0$ and $g_s=0$) and the low-symmetry case (which realizes in the low-temperature phase  of SrCu$_{2}$(BO$_{3}$)$_{2}$) separately.  
\subsection{High symmetry case}
\label{sec:phase_diag_high}
When the space group is $I4/mcm$ (which is relevant in the high temperature phase $T>T_{\text{s}}$) and the symmetry group of the  two--dimer unit cell is $\mathcal{D}_{4h}$, the intra-dimer DM interaction $D$ is absent and only the inter-dimer $D'_\perp$ DM--interaction is finite. With this type of anisotropy the $z$ component of the spin is a conserved quantity and this greatly simplifies the form of the variational ground states and of the bond-wave Hamiltonian. 

\begin{figure}[h]
\begin{center}
\includegraphics[width=8truecm]{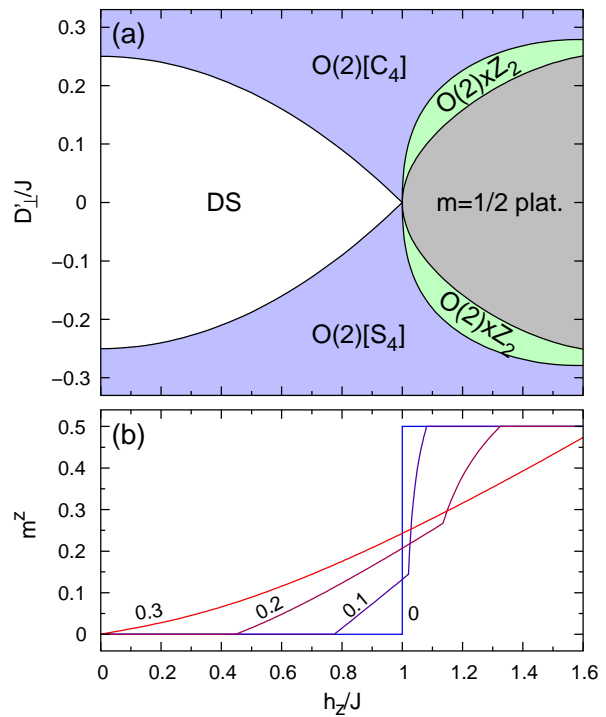}
\caption{(color online) (a) Phase diagram in the $h_z$--$D'_\perp$ plane for $D=0$ and $J'/J=0.6$. DS is the dimer singlet phase that remains a variational ground state even for finite values of $D'_\perp$. The spin configurations in the different phases are shown in Fig.~\ref{fig:phases}. `m=1/2 plat.' denotes the half--magnetization plateau phase, with a singlet and a magnetized triplet in each unit cell. (b) The magnetization curves for different values of $|D'_\perp|$ as a function of the magnetic field.} 
\label{fig:phaseD0_hz}
\end{center}
\end{figure}

Numerically minimizing the variational energy $\langle \Psi | H | \Psi \rangle/\langle \Psi | \Psi \rangle$ 
in the presence of a magnetic field along the $z$ direction, we have found three gapped phases (see Fig.~\ref{fig:phaseD0_hz}): the dimer--singlet (DS), the one--half magnetization plateau, and the fully polarized phase. Furthermore, there are four gapless phases associated with the symmetry breaking of the continuous $O(2)$ symmetry: the N\'eel, the $O(2)[\mathcal{C}_4]$, the $O(2)[\mathcal{S}_4]$, and the $O(2)\times Z_2$ phase. In these $O(2)$ phases, the $O(2)$ rotational symmetry in the $xy$ plane perpendicular to the field $h_z$ is spontaneously broken and they are the consequence of the $S^z$ being a good quantum number.  
Below we will consider the different phases and their excitations in more detail.

\subsubsection{Dimer--singlet phase}

As we mentioned earlier, the exact ground state of the SU(2) symmetric Shastry--Sutherland model  is the product of singlets on dimers: $| \psi_\text{A} \rangle = | \psi_\text{B} \rangle = |s\rangle$ for $0 \leq J' \lesssim 0.68 J$, as shown in Ref.~[\onlinecite{Koga2000}]. In the variational approach this ground states turns out to be stable for finite values of $D'_\perp$ and magnetic fields $h<h_\text{c}$, where the critical field is given by 
\begin{equation}
h_\text{c} = \sqrt{J^2-4|D'_\perp| J} \;.
\label{eq:hc}
\end{equation}
The ground state energy is coming purely from the exchange within a dimer:
 \begin{equation}
   E_\text{DS} = -\frac{3 J}{2} \; ,
   \label{eq:EDS}
 \end{equation}
all the other bond energies are identically 0.

\begin{figure}
\begin{center}
\includegraphics[width=8truecm]{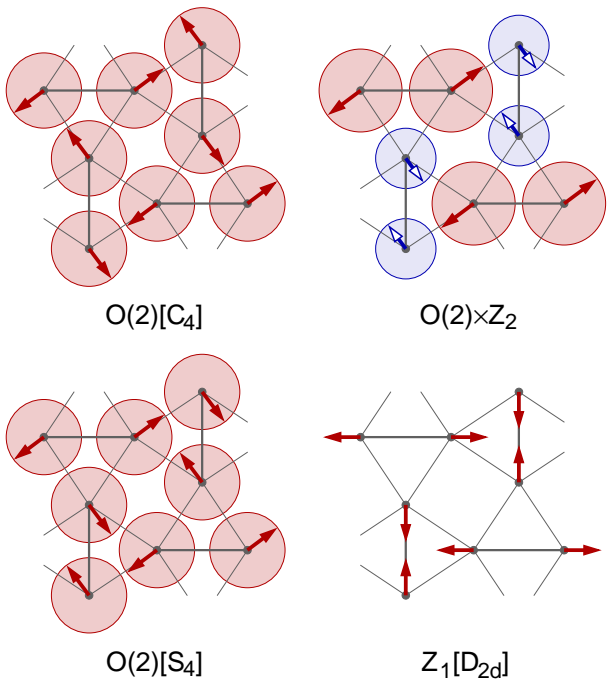}
\caption{(color online) Schematic representation of the $O(2)$ symmetric phases in the high symmetry ($D=0$) case with magnetic field perpendicular to the CuBO$_3$ layer (i.e. $h \| z$). We have plotted the expectation values of the spin component in the $xy$ plane. Since in this case $S^z$ is a conserved, the spins can be arbitrarily rotated by a global $O(2)$ rotation in the plane. The blue (open arrow) and red (solid arrow) represent inequivalent spins.}
\label{fig:phases}
\end{center}
\end{figure}

\subsubsection{The $O(2)$ phases}
\label{sec:O2}
Between the dimer singlet and the one-half magnetization plateau we find $O(2)$ symmetric phases 
(see Fig.~\ref{fig:phaseD0_hz}) when we apply the field perpendicular to the plane. In these phases the magnetization increases continuously between 0 and 1/2 per dimer (or $m^z=1$ per unit cell). Since the $S^z$ is a conserved quantity, the Hamiltonian does not break the $O(2)$ symmetry of the rotations around 
the $z$-axis. This symmetry is spontaneously broken in the $O(2)$ phase. From numerical minimization we found that the wave function can be written as 
\begin{subequations}
\label{eq:wfo2}
\begin{eqnarray}
| \psi_\text{A} \rangle &\propto& |s\rangle + u e^{i\varphi}|t_{1}\rangle + d e^{-i\varphi} |t_{\bar 1}\rangle \;, \label{eq:wfo2A}\\
 | \psi_\text{B} \rangle &\propto& |s\rangle \pm i u e^{i\varphi}|t_{1}\rangle \mp i d e^{-i\varphi}|t_{\bar 1}\rangle \;, \label{eq:wfo2B}
\end{eqnarray}
\end{subequations}
with the upper sign for $D'_\perp>0$ and the lower sign for $D'_\perp<0$. This wave function is continuously connected to the dimer--singlet phase, as setting $u=0$ and $d=0$ we get back to the product of singlets. In this phase the $S^z$ expectation values for all the spins are equal 
($\langle S^{z}\rangle \propto |u|^{2}-|d|^{2}$), and the spin components in the $xy$ plane along the inter-dimer bonds are perpendicular to each other in such a way that going around on a void square the spins also make a full turn (as shown in Fig.~\ref{fig:phases}) -- in other words, the spin configurations are invariant with respect to either the $C_4$- or the $S_4$-rotation. It is the sense of the rotation that makes the difference between the two different cases of the sign of $D'_\perp$. 
To make a clear distinction, we use the symmetry groups
which leave the variational ground state invariant to label these two phases as $O(2)[\mathcal{C}_4]$ 
and $O(2)[\mathcal{S}_4]$.  
We found that they are realized for positive and negative values of $D'_\perp$, respectively. 

 The expectation value of the Hamiltonian with the wave functions (\ref{eq:wfo2A}) and (\ref{eq:wfo2B}) reads
\begin{eqnarray}
 E_{O(2)} &=&
   \frac{u^2+d^2-3}{u^2+d^2+1} \frac{J}{2}
 +  \frac{2(u^2-d^2)^2}{\left(u^2+d^2+1\right)^2} J' \nonumber\\
&& -\frac{4(u-d)^2}{\left(u^2+d^2+1\right)^2}  |D'_\perp| 
 -   \frac{2(u^2-d^2)}{u^2+d^2+1} h_z \;, \label{eq:EO2var}
\end{eqnarray}
and the minimization gives a set of polynomial equations that needs to be solved numerically.
Close to the phase boundary to the dimer singlet phase given by Eq.~(\ref{eq:hc}), we can expand in $\delta h = h_z-h_\text{c}$. In the lowest order in $\delta h$ 
\begin{subequations}
\begin{eqnarray}
 u &=& -\frac{(J+h_{\text{c}}) \sqrt{2 h_{\text{c}}}}{2 \sqrt{4 J J' h_{\text{c}}^2 + J^4-h_{\text{c}}^4}}\sqrt{\delta h} \;, \\
 d &=& \frac{(J-h_{\text{c}}) \sqrt{2 h_{\text{c}}}}{2 \sqrt{4 J J' h_{\text{c}}^2 + J^4-h_{\text{c}}^4}} \sqrt{\delta h} \;.
\end{eqnarray}
\end{subequations}
The magnetization below $h_{\text{c}}$ is 0, and above $h_{\text{c}}$ grows  as 
\begin{equation}
m^z = 
\frac{(J-4 |D'_\perp|) \delta h}{2 J |D'_\perp| - 4 {D'_\perp}^2 + J'(J - 4 |D'_\perp|) }+O\left(\delta h^2\right)\;.
 \label{eq:mzO2}
\end{equation}

In the absence of the magnetic field the modulus of the amplitudes $u$ and $d$ of the two triplet components become equal, and writing $v/\sqrt{2}=u=-d$ the wave function simplies to 
\begin{subequations}
\begin{eqnarray}
| \psi_\text{A} \rangle &\propto& |s\rangle + \frac{v}{\sqrt{2}}  \left(e^{i\varphi}|t_{1}\rangle - e^{-i\varphi}|t_{\bar 1}\rangle \right) \label{eq:wfo2Ah0} \;,\\
 | \psi_\text{B} \rangle &\propto& |s\rangle \pm \frac{v}{\sqrt{2}}  i \left(e^{i\varphi}|t_{1}\rangle + e^{-i\varphi}|t_{\bar 1}\rangle \right) \;, \label{eq:wfo2Bh0}
\end{eqnarray}
\end{subequations}
where, as we noted,  the sign is determined by that of $D'_\perp = \pm |D'_\perp|$. The minimum of the energy Eq.~(\ref{eq:EO2var}) in this case is achieved for 
\begin{equation}
 v = \sqrt{\frac{4 |D'_\perp|-J}{4 |D'_\perp|+J}} \;,
\end{equation}
with
\begin{equation}
 E = -\frac{J}{2} - 2 |D'_\perp| - \frac{J^2}{8 |D'_\perp|} \;.
  \label{eq:EO2h0}
\end{equation}
From the analysis above it turns out that the $O(2)$ phases are realized for $|D'_\perp|> J/4$ in zero field [this is consistent with Eq.~(\ref{eq:hc})].

\subsubsection{The $O(2)\times Z_2$ phase}

 The $O(2)$ phase(s) and the one--half magnetization plateau phase are connected via two continuous phase transitions. The intermediate phase exhibits both the $Z_2$-symmetry breaking of the plateau phase 
 (the inequivalence of the $z$-component of the magnetization on the A- and the B dimers), and the $O(2)$ symmetry breaking of the $O(2)$ phase, as shown in Fig.~\ref{fig:phases}. As we approach the boundary of the one--half magnetization plateau the component of the spins perpendicular to the field decreases, and eventually vanishes at the phase boundary. Though we do not break the translational symmetry, the fact that  the magnetization along the field is not equal on the A- and the B dimer (a discrete symmetry is broken), and that at the same time we break a continuous symmetry of the $O(2)$ type, we may call this phase a supersolid.\cite{Matsuda-T-70, Liu-F-73} 
 

\begin{figure}[h]
\begin{center}
\includegraphics[width=8truecm]{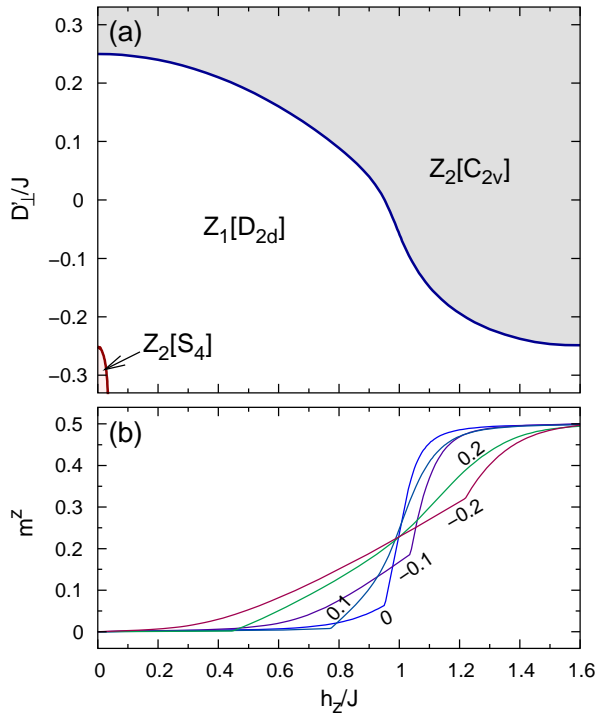}
\caption{(color online) 
(a) Phase diagram in $h_z$--$D'_\perp$ plane the for $J'/J=0.6$ and $D/J=0.1$ (the 
low--symmetry case). In comparison to the $D=0$ case in Fig.~\ref{fig:phaseD0_hz}, in a large region of the phase space the dimer-singlet and the $O(2)[S_4]$ essentially merged to create the $Z_1[D_{2d}]$ phase, a small part of the $O(2)[S_4]$ phase become a twofold degenerate $Z_2[S_4]$, and the $O(2)[C_4]$ merged with the $m=1/2$ magnetization plateau phase into  the $Z_2[C_{2v}]$ phase. (b)
Magnetization curves for a few selected values of $D'_\perp$.
\label{fig:phaseD01_hz}}
\end{center}
\end{figure}

\subsection{Low--symmetry case}

At low temperatures (specifically, $T<T_{\text{s}}=395$ [K] in SrCu$_2$(BO$_3$)$_2$), the symmetry of two--dimer unit cell is $\mathcal{D}_{2d}$. As noted earlier, the lowering of the symmetry allows for a finite value of the intra-dimer DM--interactions $D$. 
From numerical minimization, we mapped out the phase diagram, and we found a ground state that does not break any of the symmetries of the Hamiltonian in the low field region of the experimentally relevant parameters (we call the phase $Z_1[\mathcal{D}_{2d}]$) (see Fig. \ref{fig:phaseD01_hz}). 
Additionally, we found many two-fold degenerate $Z_2$ phases that we will describe in more details in a separate paper. 

The total $S^z$ is not a good quantum number any more and the continuous symmetry of the $O(2)$ phases gets reduced to discrete symmetries, and in the absence of continuous symmetry all phases become gapped.
 This symmetry reduction is seen in the expectation values of the energy;  
the inclusion of the intra-dimer DM $D$ and the staggered $g$-tensor $\tilde{g}_\text{s}$ 
introduces anisotropy terms to Eq.~(\ref{eq:EO2var}) 
\begin{eqnarray}
E_{Z_1[\mathcal{D}_{2d}]} &=& E_{O(2)} + \sqrt{2}  D\frac{ u+d}{u^2+d^2+1}\cos\varphi 
\nonumber\\ && 
- 2 \sqrt{2} \tilde g_s h_z \frac{ u-d}{u^2+d^2+1}\cos\varphi
\;,
\label{eq:EZ1var}
\end{eqnarray}
which determine the preferred direction of the $xy$ components. 
Assuming $D>0$ and positive $u$ and $v$, the DM energy on the dimers is minimal when $\varphi=\pi$. 
The variational wave function in the $Z_1[\mathcal{D}_{2d}]$ is given by
\begin{subequations}
\label{eq:WFI}
\begin{eqnarray}
| \psi_\text{A} \rangle & \propto& \left| s \right> - u \left| t_{1} \right> - d \left| t_{\bar 1} \right>  \;,
\\
| \psi_\text{B} \rangle & \propto& \left| s \right> + i u \left| t_{1} \right> - i d \left| t_{\bar 1} \right> \;, 
\end{eqnarray}
\end{subequations}
and the expectation values of spin components are shown in Fig.~\ref{fig:phases}. 
This phase is adiabatically connected to the dimer--singlet
 product phase (when $u\rightarrow 0$ and $d\rightarrow 0$) and in general the intra-dimer DM interaction 
 $D$ mixes the triplet components to the singlet, as expected from Eq.~(\ref{eq:intra}). It  is also a special case of the wave function of the $O(2)[\mathcal{S}_4]$ phase [ Eqs.~(\ref{eq:wfo2A}) and (\ref{eq:wfo2B}) with lower sign] with the phase locked to $\varphi=0$. In other words, when $D'<0$ (the experimentally relevant case) turning on an infinitesimal value of $D$ removes the phase boundary between the dimer-singlet 
 (DS) phase and the $O(2)[\mathcal{S}_4]$ phase. For $D'>0$, the $O(2)[\mathcal{C}_4]$ phase becomes frustrated with respect to $D$, and will give rise to a $Z_2$-symmetry breaking.

The conditions $\partial E_{Z_1[\mathcal{D}_{2d}]}/\partial u=0$ 
and $\partial E_{Z_1[\mathcal{D}_{2d}]}/\partial d=0$ 
lead to a set of polynomial  equations of high degree that one can solve only numerically. However, we can search for the solution of $u$ and $v$ as an expansion in $D/J$.
For small values of $D$, both $u$ and $d$ are linearly proportional to $D$ when $h_z<h_{\text{c}}$ [$h_{\text{c}}$ is defined in Eq.~(\ref{eq:hc})], and we can expand the energy as 
\begin{eqnarray}
E_{Z_1[\mathcal{D}_{2d}]} &=& -\frac{3 J}{2}
+2 J (d^2 + u^2) + 4 D'_\perp  (u-d)^2
\nonumber\\&&
-2 h_z (u^2-d^2)-\sqrt{2} D (u+d)\nonumber\\&&
 +2\sqrt{2} \tilde g_s h_z (u-v) \;,
\end{eqnarray}
 and in the lowest order in $D$  and $\tilde g_s$ the minimum is achieved with 
\begin{subequations}
\label{eq:udsmallZ1}
\begin{eqnarray}
 u &=& \frac{1}{2 \sqrt{2}} \frac{(D-2 \tilde g_s h_z)(J+h_z)+4 D D'_\perp}{J^2+4JD'_z-h_z^2} \;,\\
 d &=& \frac{1}{2 \sqrt{2}} \frac{(D+2 \tilde g_s h_z)(J-h_z)+4 D D'_\perp}{J^2+4JD'_z-h_z^2} \;.
\end{eqnarray}
\end{subequations}
For $h_z=0$ we recover Eq.~(\ref{eq:wIh0}). To be more precise, the expansion is actually in $D/(J^2+4JD'_z-h_z^2)=D/(h_{\text{c}}^2-h_z^2)$, and the denominator becomes 0 at the $D=0$, $\tilde g_s=0$ boundary between the dimer--singlet and the $O(2)[\mathcal{S}_4]$ phase.
 The expression for the magnetization of a dimer is then  
\begin{equation}
m^z = \frac{h_z}{J}\frac{h_{\text{c}}^2 }{\left(h_{\text{c}}^2-h_z^2\right)^2} (D-2\tilde g_s J)^2
    +O\left(D^4\right)  \; ,
\end{equation}
which grows quadratically with the anisotropy.
We note that $J'$ enters only in the next order in the expansion. 

\section{Bond--wave spectrum for the field parallel to $z$ axis}
In this section, we calculate the bond-wave excitation spectrum in the presence of 
an external magnetic field 
$h\parallel z$ and examine qualitatively the effect of intra and inter DM interactions.
 \subsection{High symmetry case}
 First, we consider the case where $D=0$. In the dimer--singlet product state we condense the singlet states on both the A and B bonds, and there is no need to rotate the bosons ($\tilde t = t$). The energy is then (up to a constant shift) 
\begin{equation}
 \mathcal{H} =  E_\text{DS} + \sum_{\mathbf{q} \in \text{BZ}} \mathcal{H}_2(\mathbf{q}) \;,
\end{equation}
where
\begin{widetext}
\begin{eqnarray}
\mathcal{H}_2(\mathbf{q}) & = &
J \left[
 t^{\dagger}_{0,A}({\bf q})  t^{\phantom{\dagger}}_{0,A}({\bf q})
+ t^{\dagger}_{0,B}({\bf q})  t^{\phantom{\dagger}}_{0,B}({\bf q})
\right]
\nonumber\\
&+& 
\left(
\begin{array}{l}
t^{\dagger}_{1,B}({\bf q}) \\
  t^{\dagger}_{1,A}({\bf q})\\
 t^{\phantom{\dagger}}_{\bar 1,A}(-{\bf q}) \\
 t^{\phantom{\dagger}}_{\bar 1,B}(-{\bf q}) \\
\end{array}
\right)^T \cdot \left(
\begin{array}{llll}
J - h_z  & - 2 i D'_\perp \gamma_1 & -2 i D'_\perp \gamma_1 & 0 \\
 i 2 D'_\perp \gamma_1 & J-h_z & 0 & i 2 D'_\perp \gamma_1 \\
 i 2 D'_\perp \gamma_1 & 0 & J+h_z &  i 2 D'_\perp \gamma_1 \\
 0 & -2 i D'_\perp \gamma_1 & -2 i D'_\perp \gamma_1 & J+ h_z \\
 \end{array}
\right) \cdot \left(
\begin{array}{l}
t^{\phantom{\dagger}}_{1,B}({\bf q}) \\
 t^{\phantom{\dagger}}_{1,A}({\bf q})\\
 t^{\dagger}_{\bar 1,A}(-{\bf q}) \\
 t^{\dagger}_{\bar 1,B}(-{\bf q})
\end{array}
\right)
\label{eq:HBWDS}
\end{eqnarray}
\end{widetext}
and
\begin{equation}
\gamma_1 = \cos \frac{q_a}{2} \cos \frac{q_b}{2}
 \label{eq:gamma1} \;.
\end{equation}
The Hamiltonian matrix is of the form of Eq.~(\ref{eq:Kab}), and can be diagonalized following the procedure outlined in the Appendix~\ref{sec:appendix_restricted}. The operators in the momentum space are defined by the 
$t^\dagger({\bf k}) = N_\Lambda^{-1/2} \sum_{j} e^{i {\bf k}\cdot {\bf r}_j} t^\dagger_j$, where ${\bf r_j}$ is the position of the $j$-th spin.

Actually, we can introduce the following combinations
\begin{subequations}
\label{eq:ttsym}
\begin{eqnarray}
  \tilde t^\dagger_{1,\pm}({\bf q}) &=& \frac{1}{\sqrt{2}} 
   \left[ t^\dagger_{1,A}({\bf q}) \mp i  t^\dagger_{1,B}({\bf q}) \right] \\
  \tilde t^\dagger_{\bar 1,\pm}({\bf q}) &=& \frac{1}{\sqrt{2}} 
  \left[ t^\dagger_{\bar 1,A}({\bf q}) \pm i  t^\dagger_{\bar 1,B}({\bf q}) \right] 
\end{eqnarray}
\end{subequations}
together with the corresponding annihilation operators, 
so that the original 4 by 4 matrix in Eq.~(\ref{eq:HBWDS}) decomposes into 
two 2 by 2 problems, with the Hamiltonians
\begin{eqnarray}
\mathcal{H}^{(2)}_{\pm}({\bf q})&=& 
(J-h_z\pm {D'_\perp})
 \tilde t^\dagger_{1,\pm}({\bf q})
 \tilde t^{\phantom{\dagger}}_{1,\pm}({\bf q})
\nonumber\\
&& \pm 2 {D'_\perp} \gamma_1
\left[ 
 \tilde t^\dagger_{1,\pm}({\bf q})
 \tilde t^\dagger_{\bar 1,\pm}({\bf q}) +
 \tilde t^{\phantom{\dagger}}_{1,\pm}({\bf q})
 \tilde t^{\phantom{\dagger}}_{\bar 1,\pm}({\bf q})
 \right]
\nonumber\\
&&+(J+h_z\pm {D'_\perp}) 
 \tilde t^\dagger_{\bar 1,\pm}({\bf q})
 \tilde t^{\phantom{\dagger}}_{\bar 1,\pm}({\bf q}) \;.
\end{eqnarray}

\begin{figure}[h!]
\begin{center}
\includegraphics[width=8truecm]{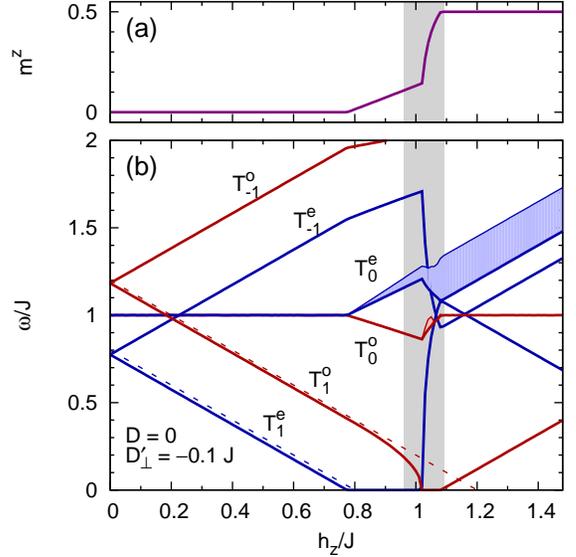}
\caption{(color online) 
Bond-wave spectrum at the $\Gamma$ point when we keep the singlets and the $S^z=1$ triplets only. The magnetic field is parallel to $z$-axis, $J'=0.6 J$, and $D'_\perp=0.1 J$. The transition into the $O(2)$ phase happens at $h_z \approx 0.77J$, and into the plateau phase at $h_z \approx 1.1 J$. For $0.77 \lesssim h_z/J \lesssim 1.03$ one of the excitations becomes identically zero --- this is the Goldstone mode of the $O(2)[\mathcal{S}_4]$ phase. The Goldstone mode of the $O(2)\times Z_2$ phase is zero for 
$1.03 \lesssim h_z/J \lesssim 1.1$.  The 2--dimer variational solution is unstable in the gray region - the dispersion goes to 0 at some wave vector away from the $\Gamma$-point at $h_z = 0.96$ and $1.08$. The filled area above the dispersion line shows the strength of the spin structure factor $S^{xx}+S^{yy}$. The dotted line is the approximation from Ref.~[\onlinecite{Miyahara2005}].
\label{fig:esr_z}}
\end{center}
\end{figure}

The bond-wave spectrum  consists of six modes: twofold degenerate nondispersing excitations with $\omega({\bf q})=J$ (denoted by $T^{\text{e,o}}_{0}$ in Fig.~\ref{fig:esr_z}) and four dispersing modes: 
\begin{eqnarray}
\omega_{+,\pm} &=& \sqrt{J^2  + 4 J {D'_\perp} \cos \frac{q_a}{2} \cos \frac{q_b}{2}} \pm h_z 
\label{eq:disphzd0even}
\end{eqnarray}
that come from $\mathcal{H}^{(2)}_{+}({\bf q})$ (blue solid lines denoted by $T^{\text{e}}_{\pm1}$ in Fig.~\ref{fig:esr_z}) and
\begin{eqnarray}
\omega_{-,\pm} &=& \sqrt{J^2  - 4 J {D'_\perp} \cos \frac{q_a}{2} \cos \frac{q_b}{2}} \pm h_z  
\label{eq:disphzd0odd}
\end{eqnarray}
from $\mathcal{H}^{(2)}_{-}({\bf q})$ that we will call the $T^{\text{o}}_{\pm1}$ modes 
(shown by red solid lines in Fig.~\ref{fig:esr_z}). The dispersions have a finite gap in the dimer-singlet product state. Let us also mention that for $h_z=0$ and $D=0$ we recover Eq.~(\ref{eq:spect}). 

 From the equations above, the gap closes at ${\bf q}=0$ when the magnetic field reaches $h_\text{c}$ defined by Eq.~(\ref{eq:hc}), and we enter into the $O(2)$ phases. Unfortunately, the explicit form of the variational wave function and the bond-wave Hamiltonian  in the $O(2)$ phases is too complicated, thus here we discuss the numerical solution only. We just mention that the combinations 
$\tilde t^\dagger_{1,\pm}({\bf q})$ and $\tilde t^\dagger_{\bar 1,\pm}({\bf q})$ introduced in Eqs.~(\ref{eq:ttsym}) are decoupling the bond-wave Hamiltonian in the $O(2)$ case as well.
When $D'_\perp>0$, the closing of the gap leads to a Goldstone mode --- the consequence of the continuous symmetry breaking ---  that appears as a continuation of the $\omega_{-,-}$ mode, and the condensation of a linear combination of the $\tilde t^\dagger_{1,-}({\bf q}=0) $ and  $\tilde t^\dagger_{\bar 1,-}({\bf q}=0) $ bosons results in the $O(2)[\mathcal{C}_4]$ phase described by the (\ref{eq:wfo2}) wave functions with the upper 
sign (see also Fig.~\ref{fig:phaseD0_hz}). For $D'_\perp<0$, on the other hand, the Goldstone mode evolves from the $\omega_{+,-}$ mode, leading to the  $O(2)[\mathcal{S}_4]$ phase (Fig.~\ref{fig:phaseD0_hz})).  
As we can see in Fig.~\ref{fig:esr_z}, the lowest gapped mode for ${\bf q}=0$ in the dimer--singlet phase remains gapless while the $O(2)$ symmetry is broken and this is the case until the half magnetization plateau. 

Also, from the ${\bf q}$-dependent excitation spectrum we learn that the spectrum may become gapless not only at the ${\bf q}=0$, but also at some other ${\bf q}$ values in the Brillouin zone, thus announcing a helical instability of the $O(2)$ phases. In Fig.~\ref{fig:esr_z} we indicate the boundary of this instability 
(the hatched region) that we have obtained from the numerical calculations of the spectra.

   
The strength of the magnetic probe response is determined by the structure factor $S^{\alpha\alpha}(\mathbf{q},\omega)$. In particular,  the $S^{xx}(\mathbf{q}=\mathbf{0},\omega)$ and $S^{yy}(\mathbf{q}{=}\mathbf{0},\omega)$ determines the strength of the ESR lines in first approximation, when the static magnetic field is along the $z$ axes. The structure factor is given by
\begin{equation}
 S^{\alpha\alpha}(\mathbf{q},\omega) 
 \propto \sum \left| \langle f | S_\mathbf{q}^{\alpha} | 0 \rangle \right|^2 
 \delta(\omega-E_f+E_0) \;,
\end{equation}
where $|0 \rangle$ is the ground state (in our case the $| \Psi \rangle $ variational wave function), by $f$ we denote the excited states, and $E_0$ and $E_f$ are the energies of the respective states. 

 As a first step, it is instructive to look at the $\omega$-integrated (static) structure factor, $S^{\alpha\alpha}(\mathbf{q}) = \int d \omega S^{\alpha\alpha}(\mathbf{q},\omega)$, which is actually the sum of the (positive) matrix elements, and that is equal to 
 $\langle \Psi | S_{-\mathbf{q}}^{\alpha}S_{\mathbf{q}}^{\alpha} | \Psi \rangle $. 
 In the (pure) dimer-singlet ground state, 
 $\lim_{{\bf q}\rightarrow 0}S_{\mathbf{q}}^{\alpha} | 0 \rangle \rightarrow 0$, 
 so we expect to see no response in ESR experiments, unless there are anisotropies 
 which mix the triplet components with the singlet. 
 
 In the $O(2)$ phase (discussed in Sec.~\ref{sec:phase_diag_high} 2), the static structure factor is  
\begin{align}
 S^{\alpha\alpha}(\mathbf{q}=\mathbf{0}) 
 &= \frac{u^2+d^2}{1+u^2+v^2} \label{eq:sumruleO2}\\
           &\approx \frac{(J^2+h_{\text{c}}^2) h_{\text{c}}}{4JJ' h_{\text{c}}^2+J^4-h_{\text{c}}^4} \delta h
\end{align}
for $\alpha = x,y$, and $z$. Examining the individual matrix elements, it turns out that the matrix elements for the $S^{xx}$ and $S^{yy}$ are all vanishing except for the $T^\text{e}_0$ line. On the other hand, the matrix elements for the $S^{zz}$ are nonzero for the $T^\text{e}_1$ and $T^\text{e}_{-1}$ lines. Since the ESR line width is proportional with $S^{xx}$ and $S^{yy}$ when the field is along the $z$-direction, we expect a strong signal for the $T^\text{e}_0$ line.

\subsection{Low symmetry case}
\begin{figure}[h!]
\begin{center}
\includegraphics[width=8truecm]{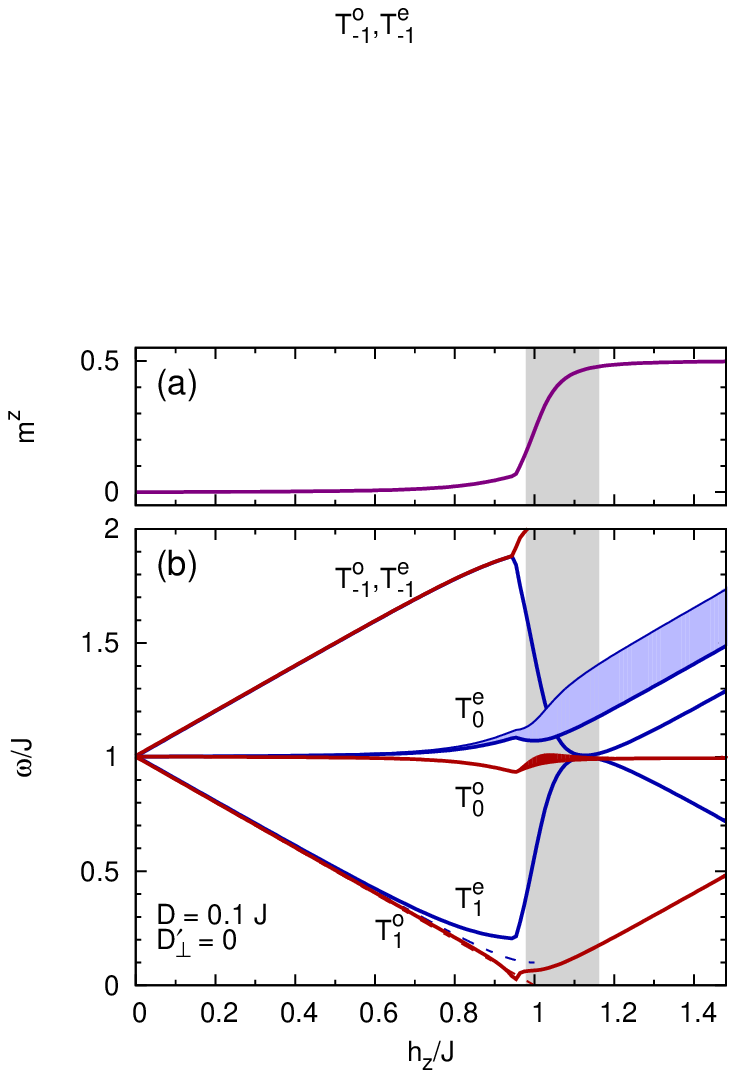}
\caption{(color online) 
Excitation spectrum in magnetic field parallel to $z$-axis when $D'_\perp=0$, $J'/J=0.6$. The instabilities toward helical states are at $h_z/J=0.9421$ and $1.1002$, while the  ${\bf k}=0$ instability into the $Z_2$ phase is at 0.9515.}
\label{fig:esr_z_0}
\end{center}
\end{figure}

Firstly let us discuss the case $D'_{\bot}=0$ shown in Fig. ~\ref{fig:esr_z_0}. 
At low fields, the spectrum looks like the usual one-triplet excitation Zeeman-split 
by the magnetic field: we see three branches that are two-fold degenerate as we have two dimers in a unit cell.
Without any kind of anisotropies, these excitations would correspond to the pure one-triplet excitations. However in the low-symmetry case the intra-dimer DM coupling $D$ mixes 
the singlets with these excitations. From the zero-field equations Eqs.~(\ref{eq:wh0}) we get that the splitting is of the order of $D^2$ for small values of $D/J$, which is much smaller than 
the linear splitting caused by $D'_{\bot}$. On the other hand, the effect of $D$ is much more pronounced at higher fields, where the gap becomes small and the singlet-triplet mixing 
is enhanced. Instead of the Goldstone mode, the anisotropy induces a ``level repulsion", and we get a finite gap that is roughly proportional to $\sqrt{D/J}$, that is consistent with 
the usual form of the anisotropy gap.  We note that the ``level repulsion" happens 
only to one of the two almost degenerate branches that come down with the applied field, and it depends crucially on the symmetry of those state as has been noted by Miyahara and Mila in Ref.~(\onlinecite{Miyahara2005}), where they considered the dispersion of a single triplet bond moving in the singlet background by the standard perturbation theory. As we increase the field, the gap closes for the $T_{1}^{\text{o}}$ level at the phase boundary to the $Z_2$ phase with $\mathcal{C}_{2v}$ symmetry (see Fig.~\ref{fig:phaseD01_hz}).

\begin{figure}[h!]
\begin{center}
\includegraphics[width=8truecm]{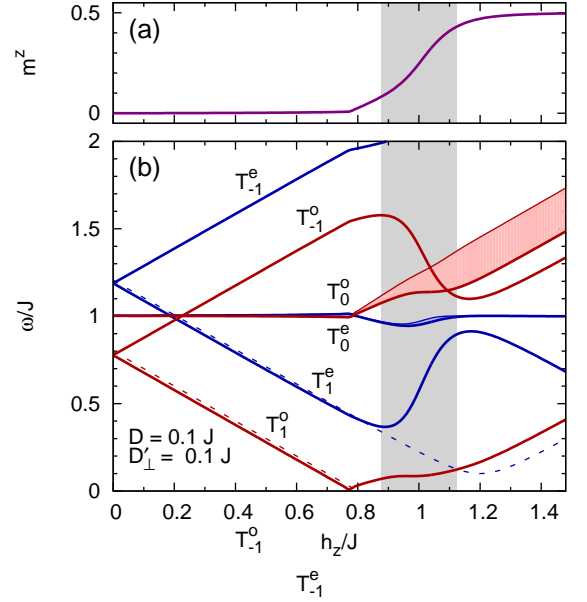}
\caption{(color online) 
Excitation spectrum in magnetic field parallel to $z$-axis in the low--symetry case, for $D'_\bot>0$, $D/J=0.1$, and $J'/J=0.6$. The notations are the same as in Fig.~\ref{fig:esr_z}.}
\label{fig:esr_z_p}
\end{center}
\end{figure}

\begin{figure}[h!]
\begin{center}
\includegraphics[width=8truecm]{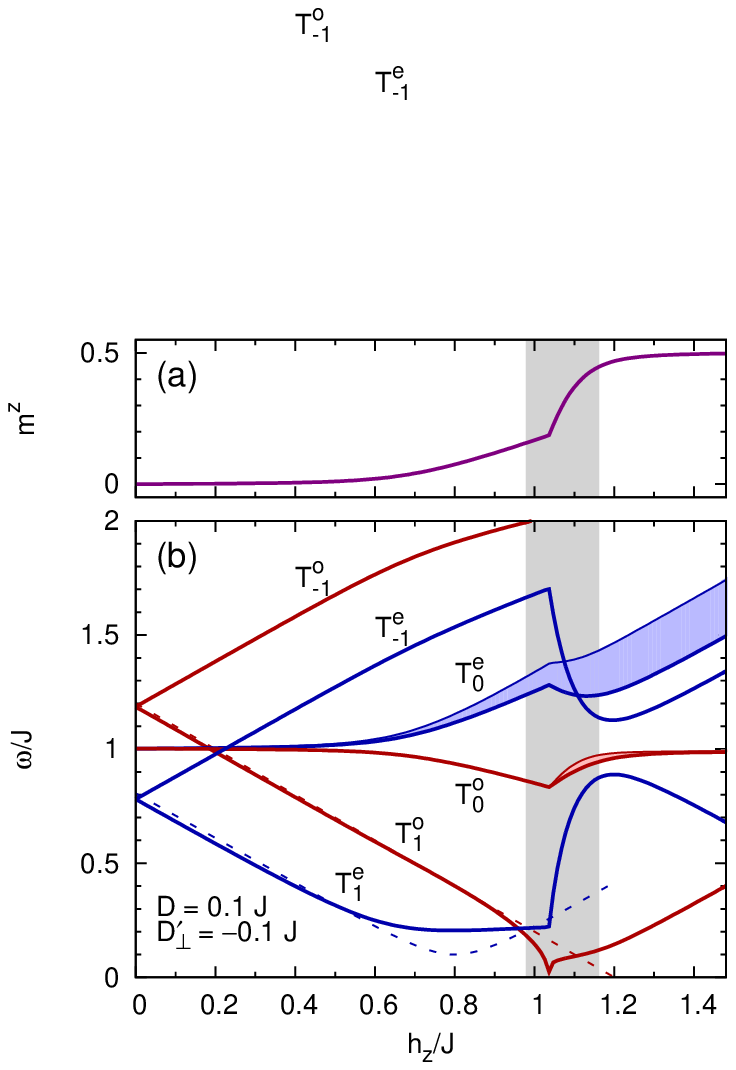}
\caption{(color online) 
Excitation spectrum in magnetic field parallel to $z$-axis for $D'_\bot<0$. ($J'=0.6 J$)}
\label{fig:esr_z_n}
\end{center}
\end{figure}

For finite inter- and intra-dimer DM interactions, we observe both the zero-field splitting and the anti level-crossing around the critical field. In Section ~\ref{sec:Dfin_h0}, we presented a detailed calculation for the zero-field dispersion and estimated this splitting in the first order of DM interactions to be $4 D'_\bot$. This is in excellent agreement with the findings of Cheng {\it et al} ~[\onlinecite{Cheng2007}]. 
The two low lying modes in Fig. ~\ref{fig:esr_z_p} and ~\ref{fig:esr_z_n}) curve differently in the $O(2)$ phase, only one of them crosses the ground state and while the other is gapped. This can be explained by that only one (the $T^e_1$) low lying excitation is coupled to $D$, and the gap is proportional to it.~[\onlinecite{Miyahara2005}]. In the case of the bond--wave calculation, the gap opens as $\sqrt{D}$ as the effect of quantum fluctuations (see Appendix \ref{sec:app_st} for more details). 
Flipping the sign of the inter-dimer DM coupling $D'_\bot$ changes the lowest-lying 
mode (compare Fig.  ~\ref{fig:esr_z_p} and ~\ref{fig:esr_z_n}).  
The singlet-triplet mixing is different according to the symmetry of the lowest-lying 
mode and the anti level-crossing occurs only for $D^{\prime}_{\perp}<0$.   

The Eq.~(\ref{eq:sumruleO2}) for the static structure factor is valid also for finite $D$ values
with the variational parameters $u$ and $v$ in the wave function (\ref{eq:WFI}) 
now being obtained by the minimization of Eq.~(\ref{eq:EZ1var}). In the limit of small $D/J$, 
we can use the values that are given by Eqs.~(\ref{eq:udsmallZ1}), and 
at ${\bf q}={\bf 0}$ we get in lowest order in $D/J$ 
\begin{equation}
 S^{xx}= S^{yy} 
  = \frac{D^2}{4} \frac{(J+4D'_\perp)^2+h_z^2}{(J^2+4JD'_{\perp}-h_z^2)^2} 
\end{equation}
for small values of field (the apparent singularity at the critical field 
$h_{z}=h_\text{c}=\sqrt{J^2-4|D'_\perp| J}$ 
is an artifact of the expansion). Similarly to the $D=0$ case, the weight of the spin correlation function $S^{xx}(\mathbf{q}=\mathbf{0},\omega)$ in the $Z_1[\mathcal{D}_{2d}]$-phase is concentrated on the single $T^\text{e}_0$ line. As we enter the $Z_2[\mathcal{C}_{2v}]$ phase, the $S^{xx}(\mathbf{q}=\mathbf{0},\omega)$ is split between the $T^\text{e}_0$ and $T^\text{o}_0$ lines. The strength of the individual lines is shown in Figs.~\ref{fig:esr_z_p}, \ref{fig:esr_z_0}, and \ref{fig:esr_z_n} by the filled region above the $T^\text{e}_0$ and $T^\text{o}_0$ lines.

\section{Phase diagram and excitation spectrum for in-plane magnetic field}
\label{sec:in-plane-field}
In this section we consider the case when the magnetic field is applied in the CuBO$_3$ plane, parallel to the bond A and perpendicular to bond B. This direction is denoted by $x$ in Fig. \ref{fig:pointgroups} (we note that this direction is different from the crystallographic $a$-axis). This choice makes the two dimers inequivalent and breaks the rotational symmetry $S_4$. This direction of field lowers the $\mathcal{D}_{2v}$ symmetry of the unit cell in the low-temperature phase to the magnetic group 
$\left\{E,\sigma_{yz}\right\}+\Theta C_2(z)\times\left\{E,\sigma_{yz}\right\}$ 
that is isomorphic to $\mathcal{C}_{2v}$.  
In the following, we show the phase diagrams for finite values of $D$ and give a short discussion 
of the phases that appear. We also show the ESR spectrum in the end of this section. 
 
 \subsection{Phase diagram}
\begin{figure}[h]
\begin{center}
\includegraphics[width=7truecm]{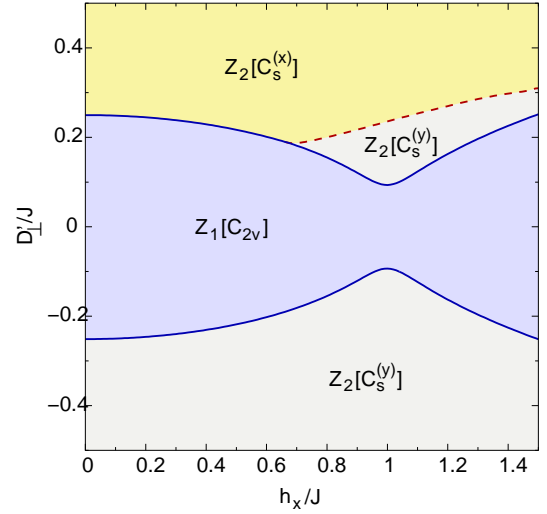}
\caption{(color online) 
Phase diagram for the magnetic field in the $x$-direction. $J'/J=0.6$ and $D/J=0.1$. The spin configurations in the different phases are shown in Fig.~\ref{fig:phases_hx}}
\label{fig:dbondD_neq_0_hx}
\end{center}
\end{figure}

\begin{figure}[h]
\begin{center}
\includegraphics[width=7truecm]{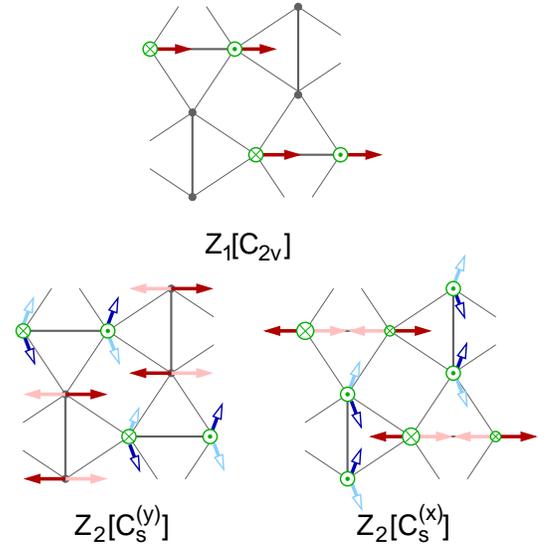}
\caption{(color online) 
Schematic representation of the spin configurations in the phases $Z_1[\mathcal{C}_{2v}]$, $Z_2[\mathcal{C}^{(x)}_s]$ and $Z_2[\mathcal{C}^{(y)}_s]$ that appear when the field is along the $x$ direction. The darker and lighter arrow represent the two degenerate states.}
\label{fig:phases_hx}
\end{center}
\end{figure}

 The numerically obtained phase diagram  as a function of $h_x$ and $D'_\perp$ for a selected value of $D$ is shown in Fig. ~\ref{fig:dbondD_neq_0_hx}. 
 We denote the ground state by $Z_1[\mathcal{C}_{2v}]$, which has the full $\mathcal{C}_{2v}$ symmetry of the Hamiltonian and the variational wave function is of the following form:
\begin{subequations}
\label{eq:gs_z1}
\begin{eqnarray}
| \psi_\text{A} \rangle &\propto&|s\rangle - v_y |t_y\rangle- i u_z |t_z\rangle \;,\\
| \psi_\text{B} \rangle &\propto&|s\rangle + v_x |t_x\rangle \;,
\end{eqnarray}
\end{subequations}
with the energy expectation value:
\begin{multline}
E_{Z_1[\mathcal{C}_{2v}]}  \\
= -\frac{J(1-v^2_x)+2 D v_x}{2(1+v^2_x)} - \frac{ J + 2 h_x u_z v_y + D u_z}{1+u^2_z+v^2_y}.\label{eq:e0_z1}
\end{multline}
 The intra-dimer DM interaction $D$ prefers states with dipole expectation values that are perpendicular to the vector $\mathbf{D}$. As a consequence, the magnetic field induces the moment only on dimers where the direction of $\mathbf{D}$ is perpendicular to the field. In our case $h||x$, and only the dimer A develops a finite magnetization: the spin components along the magnetic field, $S^x_{A,1}=S^x_{A,2}$ as well as the components perpendicular to the plane, $S^z_{A,1}=-S^z_{A,2}$ become finite, as shown in Fig.~\ref{fig:phases_hx}. Increasing the magnetic field, the expectation value of $S^x_A$ increases smoothly up to $m^x=1/2$ [see Fig. ~\ref{fig:esr_x_pm} (a)]. 
 
The wave-function $| \psi_\text{B} \rangle$, on the other hand, is time-reversal invariant, where the expectation value of any spin component is zero: $\langle \mathbf{S}_{1,2}\rangle=\mathbf{0}$. However, it breaks the rotational symmetry, as the vector chirality is finite: 
$\langle \mathbf{S}_1\times\mathbf{S}_2\rangle =-v_x/(1+v_x^2) $. This is the so called $p$-type nematic state.\cite{Andreev1984,Lauchli-D-L-S-T-05} The parameter $v_x$ does not depend on the magnetic filed, and minimizing the energy (\ref{eq:e0_z1}) we find that $v_x=D/2 J$.

 On the phase diagram, we found two additional phases, and they are both two-fold degenerate 
 $Z_2$-phases. 
For sufficiently large negative values of $D'_\bot$, the phase $Z_2[\mathcal{C}^{(y)}_s]$ is realized with the wave function
\begin{subequations}
\begin{eqnarray}
\left|\psi_\text{A} \right\rangle&\propto&\left|s\right>-(v_y \pm i u_y ) \left|t_y\right>-(i u_z\mp v_z) \left|t_z\right> \;,
\\
\left|\psi_\text{B} \right\rangle&\propto&\left|s\right>+(v_x \pm i u_x ) \left|t_x\right> \;.
\end{eqnarray}
\end{subequations}
The magnetization on the A dimer consist of a uniform part $S^x_{A,1}=S^x_{A,2}$ and a  staggered part $S^y_{A,1}=-S^y_{A_2}$ and $S^z_{A,1}=-S^z_{A,2}$ (as shown in Fig. ~\ref{fig:phases_hx}).  While in phase $Z_1[\mathcal{C}_{2v}]$ there were no dipole components on dimer $B$, here the expectation value of staggered magnetization $m^x_{B,st}$ is non-zero, $S^x_{B,1}=-S^x_{B,2}$. The magnetization pattern is invariant under $\sigma_{yz}$.

At large enough positive $D'_\bot$, we reach the phase $Z_2[\mathcal{C}^{(x)}_s]$ where the ground state has the following form
\begin{subequations}
\begin{eqnarray}
\left|\Psi\right>_{\text{A}}&=&\left|s\right>\pm i u_x\left|t_x\right> - v_y \left|t_y\right> - i u_z\left|t_z\right> \;,\\
\left|\Psi\right>_{\text{B}}&=&\left|s\right> + v_x\left|t_x\right> \pm i u_y\left|t_y\right> \pm v_z\left|t_z\right> \;.
\end{eqnarray}
\end{subequations}
In this case, the spin expectation values are, as is shown in Fig.~\ref{fig:phases_hx}, 
invariant under the reflexion $\Theta\sigma_{xz}$.

We note that in the limit of $D\rightarrow 0$ the $Z_1[\mathcal{C}_{2v}]$-phase is continuously connected to the dimer-singlet phase that is realized for 
 \begin{equation}
h_x < \sqrt{J^2- 16 {D'_\perp}^2} \; 
\end{equation}
when $D=0$. It is also continuously connected to one of the ground states of the twofold-degenerate $m=1/2$ plateau phase (where the singlets are located on the $B$ bonds).

\subsection{ESR spectra}
In the following we discuss the effect of the DM components on the ESR spectrum for the magnetic filed parallel to $x$-axis. We remind the reader that in the absence of the anisotropy (i.e. DM interactions) 
the dimer singlet is the ground state for low fields $h_x < J$ and that the excitations are the pure, 
Zeeman-split triplets with energies $J-h_x$, $J$, and $J+h_x$ each of which is twofold degenerate 
corresponding to the two dimers in the unit cell.  

\begin{figure}[h!]
\begin{center}
\includegraphics[width=7truecm]{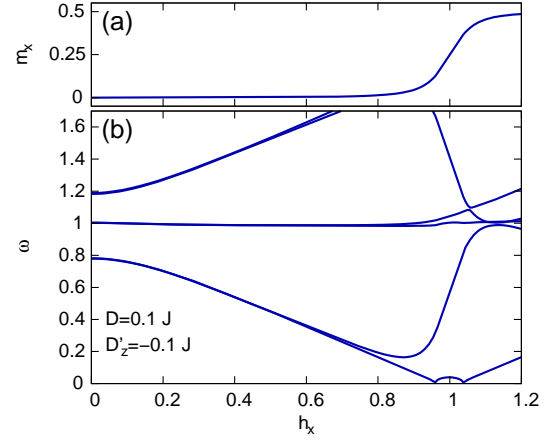}
\caption{(color online) Excitation spectrum in magnetic field parallel to $x$-axis. ($J'=0.6 J$)}
\label{fig:esr_x_pm}
\end{center}
\end{figure}

In Fig.~\ref{fig:esr_x_pm}, we show the calculated ESR spectrum for $D=0.1 J$ and $D'_\perp = -0.1 J$. In the absence of the field we observe the zero field splitting $4 D'_\perp $ that we discussed in Sec.~\ref{sec:Dfin_h0}. Now, unlike the case of the field along the $z$-direction, 
the spectrum consists of three pairs of almost degenerate levels (note that in the absence 
of $D$ each pair is exactly degenerate in the dimer singlet phase), 
and only at higher fields near the the phase transition the lines split. When $D'_\perp$ is large enough, 
the gap closes at the boundary to the $Z_2[\mathcal{C}^{(y)}_s]$-phase.  
For $D'_\perp = 0.1 J$ the spectrum looks essentially the same. 


For $D'_\bot=0$ (i.e. only $D$ is present), the zero-field splitting disappears and as we approach the critical field the two-fold degenerate triplet branch splits. The different behaviors of the two low-lying excitations around the critical field is due to their different singlet-triplet mixing. As has been discussed previously when $D$ is finite and $D'_\bot$ is zero we are in phase $Z_1[\mathcal{C}_{2v}]$ (see the phase diagram Fig.~\ref{fig:dbondD_neq_0_hx}). Increasing the magnetic field from zero, the value of the parameter $u_z$ in the ground state wave function $\left|\psi_\text{A} \right\rangle$ (Eq. ~(\ref{eq:gs_z1})) increases continuously developing a finite magnetization $m^x$ at dimer A, while the magnetization of dimer B remains zero. 
\section{Comparison to experimental spectra}
\label{sec:exp}
The ESR spectra have been considered previously by perturbation\cite{Cepas2001} 
(for $\mathbf{D}=\mathbf{0}$) 
and by exact diagonalizations\cite{Shawish2005}.  
In our approach, it is straightforward to take all the anisotropies which are relevant 
in experiments into account and  
below we consider the ESR spectra in a more realistic setting
to test our theoretical framework.

Various attempt have been made to determine the values of the different terms in the Hamiltonian. 
For completeness, we write down the Hamiltonian in its full form:
\begin{eqnarray}
\mathcal{H}&=&J\sum_{n.n.}{\bf S}_{i}{\bf S}_{j}+J'\sum_{n.n.n.}{\bf S}_{i}{\bf S}_{j}\nonumber\\
&&+\sum_{\text{n.n}} \mathbf{D}_{ij} \left(\mathbf{S}_{i}\times\mathbf{S}_{j}\right)+\sum_{\text{n.n.n}} \mathbf{D'}_{ij}\cdot\left(\mathbf{S}_{i}\times\mathbf{S}_{j}\right)\nonumber\\
&&- \sum_{u.c.}g_z h_z \left( S^{z}_{A1}+S^{z}_{A2}+S^{z}_{B1}+S^{z}_{B2} \right)  \nonumber\\
  && + \sum_{u.c.} g_s h_z \left( S^{x}_{A1}-S^{x}_{A2}+S^{y}_{B1}-S^{y}_{B2} \right) \;.
\label{eq:nojiri_model}
\end{eqnarray}
Using the estimations of C\'epas {\it et al.} (Ref.[\onlinecite{Cepas2001}]) for the value of the inter-dimer DM component $(D'_\bot/J=-0.02)$,  Kodama {\it et al} determined the intra-dimer DM component via fitting exact diagonalization data and obtained\cite{Kodama2005} $D/J=0.034$. Fitting the neutron scattering data at zero field\cite{Gaulin2004}, Cheng {\it et al.} predicted\cite{Cheng2007} $D'_{\bot}=0.18$ meV 
and $D'_{||,\text{s}}+J' D /2J=0.07$ meV. Using {\it ab initio} LSDA+U calculation, Mazurenko {\it et al.} 
estimated\cite{Mazurenko2008} $D=0.35$ meV for intra- and $D'_\bot=0.1$ meV, $D'_{||,\text{s}}=0.06$ meV and $D'_{||,\text{ns}}=0.04$ meV for inter-dimer components.

The values of the $g$-tensor anisotropies were estimated from ESR and NMR measurements: $g_x=g_y=2.05$ and $g_z=2.28$ in  Ref.~[\onlinecite{Nojiri1999}], and $g_\text{s}=0.023$ from the tilt angle of the electric field gradient in Ref.~[\onlinecite{Kodama2005}].

Information on the excitation spectra are available from ESR,\cite{Nojiri1999,Nojiri2003} Raman,\cite{Gozar2005,Lemmens2000} far-infrared (FIR) spectroscopy,\cite{Room2000,Room2004} and neutron scattering\cite{Kageyama2000,Cepas2001,Gaulin2004} measurements. We will mainly compare our spectra 
with the ESR, far-infrared, and neutron-scattering measurements.
Our lines $T^{\text{o}}_{\pm1}$ correspond to $T_{0p}(\pm)$ in the FIR spectra in Ref.~[\onlinecite{Room2004}] and to $O_1$ in the ESR spectra in Ref.~[\onlinecite{Nojiri2003}], the lines
$T^{\text{e}}_{\pm1}$ correspond to $T_{0m}(\pm)$ in [\onlinecite{Room2004}] and to $O_2$  Ref.~[\onlinecite{Nojiri2003}], and the lines $T^{\text{e}}_0$ and $T^{\text{o}}_0$ are $T_{0p,m}(0)$ 
in [\onlinecite{Room2004}].
 \subsection{Quantitative comparison to experiments at zero field}
 \label{sec:qualitative-comp}
 In particular, high resolution ESR measumerents of Nojiri {\it et al.}\cite{Nojiri2003} sees the two triplet excitations at $679\pm 2$GHz and $764\pm 2$GHz. The FIR measurements  
 of R\~o\~om {\it et al.}\cite{Room2004} observed three triplet modes at 22.72$\pm$0.05cm$^{-1}$ ($\approx$681GHz), 24.11$\pm$0.05cm$^{-1}$ ($\approx$723GHz), and at 25.51$\pm$0.05cm$^{-1}$ ($\approx$765GHz). The origin of the signal at 24.11cm$^{-1}$ is the $\Delta S_z =0$ triplet excitation that is not seen in the zero field ESR spectra. From Eq.~(\ref{eq:spect}) we can deduce 
that the splitting between the 
$\Delta S^z=1$ and $\Delta S^z=-1$ triplet lines gives $4D'_\perp\approx85$ GHz, that is $D'_\perp\approx21$ GHz.  

Furthermore, high resolution inelastic neutron scattering measurements in zero field performed 
by Gaulin {\it et al.}\cite{Gaulin2004} have revealed that the dispersion above the gap consist of three distinct branches of triplet excitations. The splitting observed there has been fitted by Cheng {\it al.}\cite{Cheng2007}  
to yield the result which is identical to our Eq.~(\ref{eq:spect}). From these dispersions, 
the splitting between the triplets at $\textbf{q}=0$ is $\Omega_{(0,0)}=4 D'_{\bot}\approx 0.4$meV (that is $\approx$95 GHz, close to the 85GHz given above), while at $\textbf{q}=(\pi,0)$ it is $\Omega_{(\pi,0)}=\sqrt{2}\left(2D'_{||,s}-\frac{D J'}{ J}\right)=0.2$meV.

We need to mention here that in our approach the dispersion of the triplets coming from the inter-dimer coupling $J'$ is altogether missing; this is why we have used, in estimating 
$D^{\prime}_{\perp}$, the `bare' value $J$ of the single-triplet 
gap, instead of using the renormalized value which is actually observed in ESR measurements 
(the above value may be renormalized if we go beyond the linear bond-wave approximation). 
Numerical diagonalization of Cheng {\it al.}\cite{Cheng2007} has shown that the effect of $J'$ is only to modify the dispersions in such a way that the splittings remain independent of the $J'$. 

\subsection{Quantitative comparison of the spectra at finite magnetic field}
In this subsection, we try to fit the ESR spectrum of Nojiri {\it et al.} in Ref.~[\onlinecite{Nojiri2003}] 
by using the bond-wave method on top of variational calculation. 
To obtain a quantitatively good fit, we need to include the DM interactions as well as the $g$-tensor anisotropies . 
\begin{figure}[h!]
\begin{center}
\includegraphics[width=8truecm]{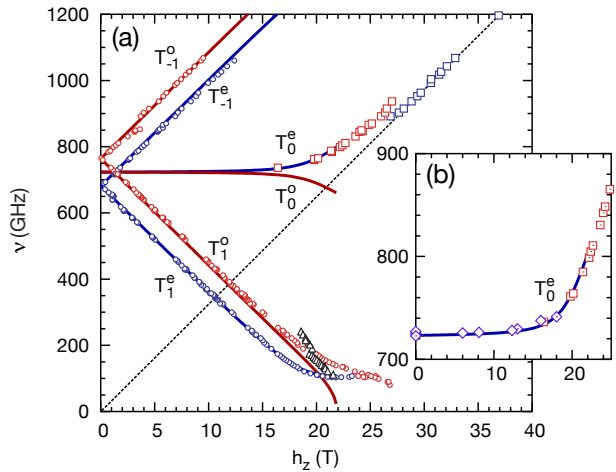}
\caption{(color online) 
Qualitative comparison of excitation spectrum with the $h \| c$ ESR spectrum shown in Fig. 4(a) in Nojiri {\it et al.}[\onlinecite{Nojiri2003}]. In the inset (b) the dimanonds are the far--infrared data from Ref.~[\onlinecite{Room2004}], and the squares are the ESR data from Ref.~[\onlinecite{Nojiri2003}]  }
\label{fig:esr_z_nojiri}
\end{center}
\end{figure}

In the fitting, we use the values of the anisotropy constants $g_z=2.28$ estimated in Ref.~[\onlinecite{Nojiri1999}], and $g_\text{s}=0.023$ in Ref.~[\onlinecite{Kodama2005}]. The value of the intra-dimer DM coupling $D=0.034 J\approx 60$ GHz is obtained in Ref.~[\onlinecite{Kodama2005}], assuming that $J=85$K (Ref.~[\onlinecite{Miyahara1999,Miyahara2000}], and a similar value ($D=1.8$ cm$^{-1}=54$ GHz) is reported in Ref.~[\onlinecite{Room2004}]. The inter-dimer coupling constant is given by $D'_\bot=21$ GHz, as determined above. 
For the reason described above, we choose $J$ to be equal to 722 GHz, the value of the experimentally observed gap\cite{Nojiri1999}. 
 Furthermore, we find that the spectrum is essentially independent of the value of $J'$ inasmuch as we are in the $Z_1[\mathcal{D}_{2d}]$ phase, so we have chosen $J'/J=0.6$ for internal consistency of the calculation. The calcualted bond--wave spectrum with the parameters mentioned above is shown in Fig.~\ref{fig:esr_z_nojiri}.

We find a surprisingly good quantitative agreement with the high-field ESR of Nojiri {\it et al} (Ref.~[\onlinecite{Nojiri2003}]) and the FIR measurements of R\~o\~om {\it et al.} (Ref.~[\onlinecite{Room2004}]). Our spectra reproduced not only the value of the high-field gap in the $T^\text{e}_1$-excatiation above 20T, but also the behavior of the $T^\text{e}_0$ level which follows nicely the main (i.e. largest-intensity) peak in the ESR spectrum [Fig.~\ref{fig:esr_z_nojiri}(b)], thus clearly identifying those lines as originating from triplet excitations.  


\section{Summary and conclusions}
We have studied the description of the magnetic properties of SrCu$_{2}$(BO$_{3}$)$_{2}$ using the Shastry--Sutherland model extended with anisotropies. The possible form of the anisotropies, like the Dzyaloshinskii-Moriya interactions [Eqs.~(\ref{eq:HD})-(\ref{eq:HDp})] and the $g$-tensor anisotropy [Eq.~(\ref{eq:gtensor})] follows from the structure and the symmetry properties (space group) of the material. 
 
 We used a bond-factorized form of the variational wave function to study the effect of the anisotropies on the ground state properties in phases which are compatible with the crystallographic unit cell comprising of two orthogonal dimers in the presence of external magnetic field. This includes the experimentally relevant dimer--singlet phase in low fields below the 1/8 magnetization plateau and the 1/2 plateau.  
  
  We have found that the dimer-singlet phase remains a good variational ground state in the so-called high temperature phase, where the only anisotropy, that takes on a finite but small value, is the inter-dimer interaction $D'_\perp$ perpendicular to the CuBO$_3$ planes, and the magnetic field is perpendicular to the plane. This phase is surrounded by gapless phases with an $O(2)$ symmetry where, due to $D'_\perp$, the triplons can propagate with a well-defined dispersion.
  
  In the less symmetrical, low-temperature structure of SrCu$_{2}$(BO$_{3}$)$_{2}$, the finite intra-dimer DM interaction $D$ appears and it gives rise to admixture of triplet components and the singlet states in the variational wave function of the `dimer-singlet' phase, without any symmetry breaking (we denoted this phase as $Z_1[\mathcal{D}_{2v}]$). 
The finite $D$ not only removes the $O(2)$ degeneracy in the gapless phases, but also introduces frustration depending on the sign of $D'_\perp$; in the unfrustrated case there is a crossover from the $O(2)$ phase 
to $Z_{1}[{\cal D}_{2d}]$ where there is a preferred direction in the $O(2)$-plane set by $D$, 
while in the frustrated case an Ising like $Z_2$-symmetry breaking appears. This is a similar behavior that has been seen in ladders\cite{DMladder} and square antiferromagnets\cite{Sato2004} with DM interactions.
  
 We also studied the effect of the anisotropies on the excitation spectra. For that purpose, we have used the ``bond-wave" theory, that is based on the bosons representing the entangled states of the dimers. In zero field we have recovered the momentum-dependent splitting of the triplet states, in accordance 
 with the neutron-scattering experiments. Furthermore, we have also recovered the experimentally measured  ESR spectra for a physically reasonable set of parameters. In this respect, we note the followings when comparing to the usual perturbational approach, where a single triplet excitation propagates\cite{Miyahara2005}: (i) in order to describe the spectra in the field that is parallel to the CuBO$_3$-plane the inclusion of all the triplet states is needed; (ii) the finite-field gap in the perturbational approach is proportional to the $D$, while in the bond wave approach it is proportional to $\sqrt{D}$, thus a smaller value of $D$ can already lead to observable effects.
 
 Regarding the ESR line intensities, we have found that in the high-symmetry case, the weight 
 in the spin-structure factor appears in the $O(2)$ phases and is concentrated on the line that is split from the $S^z=0$ triplet excitation, and, loosely speaking, follows the ``paramagnetic" ESR line. In the low-symmetry case, the anisotropies make the dimer singlet--$O(2)$ quantum critical point a crossover, and the weight  accordingly appears at the energies of the order of the zero-field singlet-triplet gap. This large weight is clearly observed in the ESR spectra in the 15-25T range at 700GHz and above.
 
 Finally, let us mention the main disadvantage of the bond-wave method. Namely, keeping only the quadratic terms in the boson operators, the dispersion due to $J'$ is not taken into account, and similarly, the $J'$ term does not decrease the singlet-triplet gap from its bare value $\Delta = J$ (for that we need to go to higher orders in the $1/M$ expansion and to introduce terms with four bosons). In other words, the situation is in this respect similar to the perturbational approaches that start from the decoupled dimers. 

\begin{acknowledgments}
We are pleased to thank  D. H\"uvonen, G. Kriza,  F.~Mila, S.~Miyahara, U. Nagel, T. R\~o\~om, and M. Zhitomirsky for stimulating discussions and M.~Hagiwara and S.~Kimura for sharing their 
unpublished ESR results and discussions. We also thank the authors of Ref.~[\onlinecite{Room2004}] for sending us their experimental data.
This work was supported under the Grant-in-Aids for Scientific Research (C) 20540375 from MEXT, Japan, by the global COE program `The next generation of physics, spun from universality and emergence' of Kyoto University, and by the Hungarian OTKA K73455 and NN76727. The authors are also thankful for the hospitality of the Yukawa Institute in Kyoto, where this work has been initiated, National Institute of Chemical Physics and Biophysics in Tallinn, where we learned about the far--infrared measurements, and MPI Physik Komplexer Systeme in Dresden, where this work has been finalized.
\end{acknowledgments}
\appendix
\section{bond-wave theory}
\label{sec:Appendix_bondwave}
 \subsection{General case}
 \label{sec:appendix_general}
While it is a welknown procedure, for completeness we present how to diagonalize a quadratic form of bosonic operators of the form
\begin{equation}
  \mathcal{K} = \frac{1}{2} 
\left(
\begin{array}{c}
  \mathbf{a}   \\
  \mathbf{a^\dagger}   
\end{array}
 \right)^T
\left(
\begin{array}{cc}
 B & A   \\
 A^\dagger & B^T 
\end{array}
 \right)
\left(
\begin{array}{c}
   \mathbf{a^\dagger}  \\
  \mathbf{a}   
\end{array}
 \right) \;,
 \label{eq:Kaa}
\end{equation}
where $B=B^\dagger$ is a hermitian $d\times d$ matrix and $A=A^T$ is a symmetric $d\times d$ matrix, so that the whole Hamiltonian $\mathcal{K}$ is hermitian (however, it is not normal ordered). The 
$\mathbf{a^\dagger}$ denotes a vector of $d$ bosons $(a^\dagger_1,a^\dagger_2,\dots,a^\dagger_d)$, similarly 
$\mathbf{a} = (a_1,a_2,\dots,a_d)$. We also assume that the $2d\times2d$ matrix in Eq.(\ref{eq:Kaa}) is positive definite -- this ensures that all the eigenvalues associated with creation operators are positive.

Applying the
$ \partial \mathcal{O}/\partial t = i\left[\mathcal{K},\mathcal{O}\right] $ 
equation of motion of an operator $\mathcal{O}$ to the bosonic operators $\mathbf{a^\dagger}$ and $\mathbf{a}$ we get:
\begin{eqnarray}
\frac{\partial}{\partial t}
\left(
\begin{array}{c}
   \mathbf{a^\dagger}  \\
  \mathbf{a}   
\end{array}
 \right) 
= i
\left(
\begin{array}{cc}
 B & A   \\
 -A^\dagger & -B^T 
\end{array}
 \right)
\left(
\begin{array}{c}
   \mathbf{a^\dagger}  \\
  \mathbf{a}   
\end{array}
 \right) \;.
\label{eq:motion}
\end{eqnarray}
Our aim is to find a suitable linear combination of $\mathbf{a}$ and $\mathbf{a^\dagger}$ operators that are energy eigenstates of the $\mathcal{K}$. This is achieved by solving the 
\begin{equation}
\left(
\begin{array}{cc}
 B & A   \\
 -A^\dagger & -B^T 
\end{array}
 \right)^T
\left(
\begin{array}{c}
   \mathbf{u}_j  \\
  \mathbf{v}_j   
\end{array}
 \right)
 = \omega_j 
 \left(
\begin{array}{c}
   \mathbf{u}_j  \\
  \mathbf{v}_J   
\end{array}
 \right) 
 \label{eq:motion2}
\end{equation}
 eigenvalue equation. The equation above has a particular property: for each $(\mathbf{u}_j,\mathbf{v}_j)$ eigenvector with eigenvalue $\omega_j>0$ the  $(\mathbf{v}^*_j,\mathbf{u}^*_j)$ is also an eigenvector with eigenvalue $-\omega_j$. We associate the eigenvectors with positive eigenvalues with creation, and with negative eigenvalues with the corresponding annihilation operator:
\begin{subequations}
\begin{eqnarray}
  \alpha^\dagger_j &=& \mathbf{u}_j\cdot\mathbf{a^\dagger} + \mathbf{v}_j \cdot \mathbf{a} \;,\\
  \alpha_j &=&  \mathbf{v}^*_j\cdot\mathbf{a^\dagger} + \mathbf{u}^*_j\cdot\mathbf{a} \;,
\end{eqnarray}
\end{subequations}
so that the 
$[\alpha_j,\alpha^\dagger_{j'}]=\delta_{j,j'}$ 
commutation relation is fulfilled. With this choice $[\mathcal{K},\alpha^\dagger_j] = \omega_j \alpha^\dagger_j$ and $[\mathcal{K},\alpha_j] = -\omega_j \alpha_j$ holds and $\mathcal{K}$ takes the 
\begin{equation}
  \mathcal{K} = \sum_{j=1}^{d} \omega_j \left( \alpha^\dagger_j \alpha_j + \frac{1}{2} \right) 
\end{equation}
diagonal form.

 \subsection{Restricted case}
\label{sec:appendix_restricted}
In certain cases the quadratic term is simpler and can be written as
\begin{equation}
  \mathcal{K'} = \left( \begin{array}{c}
       \mathbf{a}   \\
       \mathbf{b^\dagger}   
     \end{array} \right)^T
\left( \begin{array}{cc}
 B & A   \\
 A^\dagger & C  
\end{array} \right)
\left(
\begin{array}{c}
   \mathbf{a^\dagger}  \\
  \mathbf{b}   
\end{array} \right) 
\label{eq:Kab}
\end{equation}
where the $\mathbf{a^\dagger}=(a^\dagger_1,a^\dagger_2,\dots,a^\dagger_{d'})$ and  
$\mathbf{b} = (b_1,b_2,\dots,b_{d''})$, with $d=d'+d''$, and the $a$ and $b$ bosons commute among each other.This happens frequently for the bosons in the momentum representation, as the momentum conservation allows only terms of the type 
$a^\dagger_\mathbf{k}a^{\phantom{\dagger}}_\mathbf{k}$, $a^\dagger_\mathbf{-k}a^{\phantom{\dagger}}_\mathbf{-k}$, $a^\dagger_\mathbf{k}a^{\dagger}_\mathbf{-k}$, and $a^{\phantom{\dagger}}_\mathbf{k}a^{\phantom{\dagger}}_\mathbf{-k}$ (i.e. terms like $a^\dagger_\mathbf{k}a^{\phantom{\dagger}}_\mathbf{-k}$ and $a^{\phantom{\dagger}}_\mathbf{k}a^{\phantom{\dagger}}_\mathbf{k}$ are missing), and we can associate the $a_\mathbf{k}$ and $a_\mathbf{-k}$ bosons with $a$ and $b$ bosons in Eq.~(\ref{eq:Kab}), respectively. The $B=B^\dagger$ and $C=C^\dagger$ ensures that the whole Hamiltonian matrix is hermitian, moreover we assume it to be positive semidefinit.

Using the equation of motion technique, we get the equivalent of the Eq.~(\ref{eq:motion2}) for this case:
\begin{equation}
\left(
\begin{array}{cc}
 B & A   \\
 -A^\dagger & -C  
\end{array}
 \right)^T
\left(
\begin{array}{c}
   \mathbf{u}_j  \\
  \mathbf{v}_j   
\end{array}
 \right)
 = \omega'_j 
 \left(
\begin{array}{c}
   \mathbf{u}_j  \\
  \mathbf{v}_j   
\end{array}
 \right) \;.
\end{equation}
We define the creation operators and energies as
\begin{equation}
\begin{array}{ccc}
\alpha^\dagger_j = \mathbf{u}_j\cdot\mathbf{a^\dagger} + \mathbf{v}_j\cdot\mathbf{b}  & \text{and} &  \omega_j=\omega'_j   
\end{array}
\end{equation}
when $\omega'_j \ge 0$, and 
\begin{equation}
\begin{array}{lcc}
   \alpha^\dagger_j = \mathbf{v}^*_j\cdot\mathbf{b^\dagger} + \mathbf{u}^*_j\cdot\mathbf{a}  
   & \text{and}  
   & \omega_j=-\omega'_j 
\end{array}
\end{equation}
if $\omega'_j < 0$, so that $[\alpha_j,\alpha^\dagger_{j'}]=\delta_{j,j'}$ the commutation relations are satisfied. 
Eventually we arrive at the 
 \begin{equation}
\mathcal{K'} = \sum_{j=1}^{d} \omega_j \left(\alpha^\dagger_j \alpha_j + \frac{1}{2}\right)+ \frac{1}{2}\left(\mathrm{Tr}\,B - \mathrm{Tr}\,C\right)
\end{equation}
diagonal form. 

\section{Keeping $|s\rangle$ and $|t_1\rangle$ only}

\label{sec:app_st}
In the case of the field parallel to the $z$ axis it is a usual practice to keep only the low lying singlet and $S^z=1$ triplet (the component aligned with the field) state of a bond. Here we are interested in the behavior of the gap close to the critical field. For that reason, we restrict the discussion to the dimer--singlet and the $O(2)[S_4]$ phase for $D=0$ and the  $Z_1[\mathcal{D}_{2d}]$ phase in finite $D$ case. 

As a first step, we define the following rotated boson operators
\begin{subequations}
\begin{eqnarray}
\tilde{s}^{\dagger}_{\text{A}}({\bf k}) &=& \cos\frac{\alpha}{2} s^{\dagger}_{\text{A}}({\bf k})  
+\sin\frac{\alpha}{2}e^{i\varphi} t^{\dagger}_{\text{1,A}}({\bf k}) \;,\\ 
\tilde{t}^{\dagger}_{\text{A}}({\bf k}) &=& \sin\frac{\alpha}{2} s^{\dagger}_{\text{A}}({\bf k})  -\cos\frac{\alpha}{2}e^{i\varphi} t^{\dagger}_{\text{1,A}}({\bf k}) \;,\\ 
\tilde{s}^{\dagger}_{\text{B}}({\bf k}) &=& \cos\frac{\alpha}{2} s^{\dagger}_{\text{B}}({\bf k}) 
 - i \sin\frac{\alpha}{2}e^{i\varphi} t^{\dagger}_{\text{1,B}}({\bf k}) \;, \\
\tilde{t}^{\dagger}_{\text{B}}({\bf k}) &=& \sin\frac{\alpha}{2} s^{\dagger}_{\text{B}}({\bf k})  
- i \cos\frac{\alpha}{2}e^{i\varphi} t^{\dagger}_{\text{1,B}}({\bf k}) \;,
\end{eqnarray}
\end{subequations}
so that the variational wave--function that comprises the above mentioned phases is given by $\left|\Psi\right>_{\text{A}}= \tilde{s}^{\dagger}_{A} \left|0\right>$ and 
$\left|\Psi\right>_{\text{B}}= \tilde{s}^{\dagger}_{B} \left|0\right>$, and we fix the phase $\varphi=0$ for convenience [see also Eq.~(\ref{eq:wfo2}) for comparison].
The expectation value of the Hamiltonian is then given by
\begin{eqnarray}
E_{0}&=& - J \left(\frac{1}{2}+\cos\alpha\right)
+ \frac{J'}{2} (1-\cos\alpha)^2 + D'_\perp \sin^2\alpha \nonumber\\ 
&&+ \frac{1}{\sqrt{2}} \tilde D \sin\alpha - g_z h_z(1-\cos \alpha) \;.
 \label{eq:Hexpect_st}
\end{eqnarray}
Here we introduce the $\tilde D= D - 2 g_s h_z$, as in this section the $D$ and the $g_s$ appear in this combination only. Minimization procedure involves solving a quartic polynomial equation that is tedious. Instead, we concentrate on the case when the anisotropy terms $\tilde D$ is small. 

We also need the bond--wave Hamiltonian. For that we introduce
\begin{equation}
 \tilde t^{\dagger}_{\pm}(\mathbf{k})  = \frac{1}{\sqrt{2}} 
 \left[ \tilde t^{\dagger}_{A}(\mathbf{k}) \pm  \tilde t^{\dagger}_{B}(\mathbf{k}) \right]
 \label{eq:tsympm}
\end{equation}
symmetric and antisymmetric combination of the rotated triplet operators that reduce the size of the matrices in the Hamiltonian.
Expanding in powers of $M$, we get 
$\mathcal{H} = E_0 M^2 + M \mathcal{H}_2 + \dots$, where we omitted higher order terms in $1/M$. The bond wave Hamiltonian $\mathcal{H}^{(2)} = \mathcal{H}^{(2)}_+ +\mathcal{H}^{(2)}_- $ is given as 
\begin{widetext}
\begin{eqnarray}
\mathcal{H}^{(2)}_\pm&=& \sum_{\mathbf{k}} \left(\begin{array}{c}
\tilde t^{\dagger}_{\pm}(\mathbf{k}) \\
\tilde t^{\phantom{\dagger}}_{\pm}(-\mathbf{k})
\end{array}\right)^T\left(\begin{array}{cc}
a \pm (b+2D'_\perp) \gamma_1&    b \gamma_1 \\
b \gamma_1  & a \pm (b+2D'_\perp) \gamma_1
\end{array}\right)\left(\begin{array}{c}
\tilde t^{\phantom{\dagger}}_{\pm}(\mathbf{k}) \\
\tilde t^{\dagger}_{\pm}(-\mathbf{k})\end{array}
\right) \;,
 \label{eq:H2pm}
\end{eqnarray}
\end{widetext}
where the $a$ can be conveniently expressed as 
\begin{equation}
 a = b -E_0  - h_z - \frac{J}{2}
\end{equation}
 and
\begin{equation}
 b = \frac{1}{2}(J'-2D'_\perp) \sin^2 \alpha \;,
 \label{eq:bdef}
\end{equation}
while
$\gamma_1$ is defined in Eq.~(\ref{eq:gamma1}).
 The Bogoliubov transformation yields the
\begin{equation}
 \omega_ \pm = \sqrt{(a \pm 2D'_\perp \gamma_1)
 (a \pm 2 b \gamma_1 \pm 2D'_\perp \gamma_1)} \;.
  \label{eq:w2kdep}
\end{equation}

%
%

\begin{figure}[bt]
\begin{center}
\includegraphics[width=8truecm]{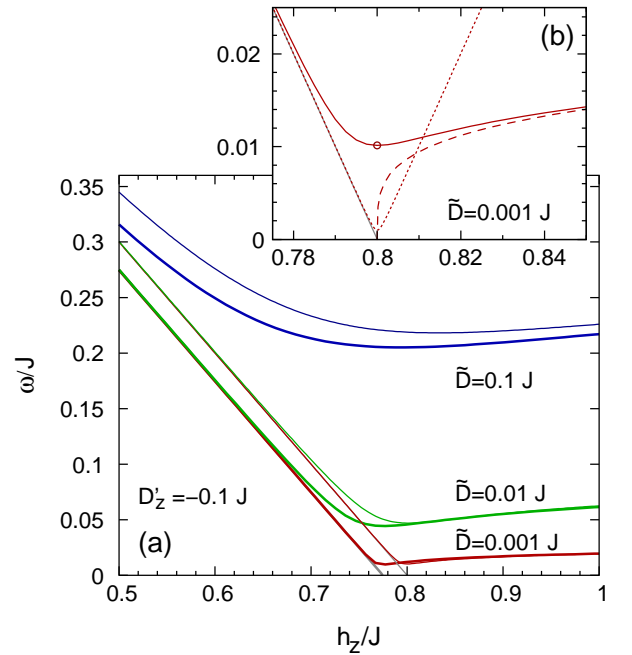}
\caption{(a) The lowest lying branch of the excitation spectrum at the $\mathbf{k}=0$ is shown when keeping 4 (thick) and 2 bosons (thin lines) per dimer for $J' = 0.6 J$, $D'_\perp = 0.1 J$ and different values of $\tilde D$. (b) The bond wave spectrum has a dip at the $h_{c1}=0.8 J$ critical field. The dotted line is the approximation from Ref.~[\onlinecite{Miyahara2005}], the dashed line and the circle are the approximations given by Eqs.~(\ref{eq:gapd1o2}) and (\ref{eq:gapd2o3}), respectively.}
\label{fig:esr_appendix}
\end{center}
\end{figure}

\subsection{High symmetry case}

The minimal energy for $\tilde D=0$ of Eq.~(\ref{eq:Hexpect_st}) is achieved for
\begin{equation}
\cos \alpha_{O(2)} = \left\{\begin{array}{ccc}
 1  &   &  h_z \le h_{c1} \;, \\
 \frac{h_{\text{c1}}+h_{\text{c2}}-2h_z}{h_{\text{c2}}-h_{\text{c1}}}&   
 & h_{\text{c1}} \le h_z \le h_{\text{c2}} \;, \\
-1   &  & h_{\text{c2}} \le h_z \;.   
\end{array}
\right. 
\label{eq:alphaO2}
\end{equation}
This solutions correspond the the dimer--singlet, $O(2)[S_4]$ and the fully polarized phase, respectively. Note that the one--half magnetization plateau is missing -- the form of the chosen wave function does not allow for the $Z_2$ breaking.
\begin{subequations}
\begin{eqnarray}
 h_{\text{c1}} &=& J-2|D'_\perp| \;,\\
 h_{\text{c2}} &=& J+2J'+2|D'_\perp| \;.
\end{eqnarray}
\end{subequations}

The variational energy of the unit cell is then
\begin{equation}
E_{0}= \left\{\begin{array}{ccc}
-\frac{3J}{2}  &   &  h_z \le h_{\text{c1}}  \\
-\frac{3J}{2} - \frac{(h_z-h_{\text{c1}})^2}{h_{\text{c2}}-h_{c1}}  &   
& h_{\text{c1}} \le h_z \le h_{\text{c2}}  \\
\frac{J}{2} + 2J' -2 g_z h_z  &   &  h_{\text{c2}} \le h_z 
\end{array}
\right. \;.
\end{equation}

 It turns out that the boundary between the dimer singlet phase and the $O(2)$ phase is shifted to the expense of the $O(2)$ phase compared to the case when we keep all the four state of a dimer [see Eq.~(\ref{eq:hc})], and the boundaries overlap only in the limit of small $|D'_\perp|$ values, when the critical field is close to $J$.

Now, let us turn to the excitation spectrum.
In the dimer--singlet phase $a = J - h_z$ and  $b=0$ in Eq.~(\ref{eq:H2pm}), so that the $\mathcal{H}^{(2)}_\pm$ matrices are actually diagonal,
\begin{eqnarray}
\mathcal{H}^{(2)}_\pm&=& 
  \sum_{\mathbf{k}}   
  \omega_\pm(\mathbf{k}) \tilde t^{\dagger}_{\pm}(\mathbf{k}) \tilde t^{\phantom{\dagger}}_{\pm}(\mathbf{k}) 
\;,
\end{eqnarray}
with the excitation energies 
\begin{equation}
\omega_\pm({\bf k})= J -h_z \pm 2 D'_\perp\cos\frac{ q_a }{2}\cos\frac{ q_b }{2} \;.
\end{equation}
This is the same as the small $D'_\perp/J$ limit of the dispersions given by Eqs.~(\ref{eq:disphzd0even}) and (\ref{eq:disphzd0odd}), when we kept all the four bosons per dimer. 

We can also write the bond--wave Hamiltonian in the 
\begin{equation}
\mathcal{H}^{(2)}= \sum_{\mathbf{k}} \begin{pmatrix}
\tilde t^{\dagger}_{A}(\mathbf{k}) \\
\tilde t^{\dagger}_{B}(\mathbf{k})
\end{pmatrix}^T
\begin{pmatrix}
J-h_z & 2 D'_{\bot} \gamma_1 \\
2 D'_{\bot} \gamma_1& J-h_z 
\end{pmatrix}
\begin{pmatrix}
\tilde t^{\phantom{\dagger}}_{A}(\mathbf{k}) \\
\tilde t^{\phantom{\dagger}}_{B}(\mathbf{k})\end{pmatrix}  \;,
\end{equation}
form: here we recognize, up to phase factors, the upper left corner of the 4 by 4 matrix in Eq.~(\ref{eq:HBWDS}).


In the $O(2)[S_4]$ phase the $a=J-h_{\text{c1}} = -2 D'_\perp$ and 
\begin{equation}
 b = \frac{(h_z-h_{\text{c1}})(h_{\text{c2}}-h_z)}{(h_{\text{c2}}-h_{\text{c1}})} \;,
 \label{eq:bO2}
\end{equation}
and from Eq.~(\ref{eq:w2kdep}) we get
\begin{eqnarray}
\omega^\pm({\bf k}) &=& 2
  \sqrt{1\mp\cos\frac{{\bf a}}{2}\cos\frac{{\bf b}}{2}}
    \nonumber\\
    &&\times \sqrt{
    D'_\perp \left(D'_\perp \mp  
             \left(D'_\perp +b \right)            
             \cos\frac{{\bf a}}{2}\cos\frac{{\bf b}}{2}  \right)} \;.
\end{eqnarray}
We note that $\omega^+({\bf k})\rightarrow 0$ as $\mathbf{k} \rightarrow 0$, thus it becomes the  Goldstone mode associated with the continuous symmetry breaking in the $O(2)$ phase.



\subsection{Low symmetry case}

In the presence of the $\tilde D$ anisotropies the $\omega_+$ Goldstone mode acquires a finite gap in the presence of anisotropies.
In the case of small $\tilde D$ we include the first order correction in $\tilde D$ to the $\alpha$ given by Eq.~(\ref{eq:alphaO2}),
\begin{equation}
\cos \alpha = 
 \left(1+ \frac{\tilde D}{\sqrt{2(h_z-h_{\text{c1}})(h_{\text{c2}}-h_z)}}\right)
 \cos \alpha_{O(2)} \;,
\label{eq:alphaO2pert}
\end{equation}
and we end up with
\begin{equation}
\omega_{+}=\tilde D^{1/2} \left[ \frac{(h_z-h_{\text{c1}})(h_{\text{c2}}-h_z)}{2}\right]^{1/4}
\label{eq:gapd1o2}
\end{equation}
in the leading order in $\tilde D$. This approximation is shown with dotted line in Fig.~\ref{fig:esr_appendix}(b). It clearly fails as $h \rightarrow h_{c1}$, as in the limit $\alpha_{O(2)} \rightarrow 0$ the Eq.~(\ref{eq:alphaO2pert}) is not valid any more.
Instead, at the critical field $h_{c1}$ and in the $\alpha \rightarrow 0$ limit the energy expression Eq.~(\ref{eq:Hexpect_st}) simplifies considerably, in leading order 
\begin{equation}
  \alpha = - \frac{2^{1/6} \tilde D^{1/3}}{(J'-2 D'_\perp)^{1/3}}\;,
\end{equation}
and for the gap we get
\begin{equation}
\omega_{\text{c1}} =  \frac{\sqrt{3} \tilde D^{2/3} \left(J'-2 D'_\perp\right)^{1/3}}{2^{2/3}} \;.
\label{eq:gapd2o3}
\end{equation}
This approximation is shown with circle in Fig.~\ref{fig:esr_appendix}(b).
We find that on the boundary between the dimer--singlet and the $O(2)$ phase the gap closes faster than $\tilde D^{1/2}$, namely with a power $2/3$. Such a behavior at the quantum critical point has been discussed for quantum antiferromagnets in Refs.~[\onlinecite{Fouet2004}] and [\onlinecite{Chernyshev2005}]. We also note that the perturbational $\sqrt{(h-h_{\text{c1}})^2+\tilde D^2}$  result of Ref.~[\onlinecite{Miyahara2005}] does not capture the quantum fluctuation effects close to the critical field $h_{\text{c1}}$.
\end{document}